\documentclass[aps,prd,onecolumn,secnumarabic,amssymb,nobibnotes]{revtex4-1}
\usepackage{graphicx}% Include figure files
\usepackage{dcolumn}% Align table columns on decimal point
\usepackage{bm}% bold math
\usepackage{color}
\usepackage{longtable}
\usepackage{supertabular}
\usepackage{appendix}
\usepackage[T1]{fontenc}

\makeatletter

\newcommand{\Rmnum}[1]{\expandafter\@slowromancap\romannumeral #1@}
\makeatother

\begin{document}

\title{Constraining {Light Scalar Field} with Torsion-Balance Gravity Experiments }

\author{Cheng-Gang Qin$^{1}$}
\author{Xiao-Yu Lu$^2$}\email[E-mail:]{xiaoyulu1993@163.com}
\author{Bing-Chen Zhao$^3$, Jun Ke$^{3}$, An-Bin Du$^{4}$}
\author{Jie Luo$^{3}$} \email[E-mail:]{luojiethanks@126.com}
\author{Yu-Jie Tan$^{1}$} \email[E-mail:]{yjtan@hust.edu.cn}
\author{Cheng-Gang Shao$^{1}$} \email[E-mail:]{cgshao@hust.edu.cn}

\affiliation
{$^1$ MOE Key Laboratory of Fundamental Physical Quantities
Measurement $\&$ Hubei Key Laboratory of Gravitation and Quantum Physics, PGMF and School of Physics, Huazhong University of Science and Technology, Wuhan 430074, People's Republic of China\\
$^2$Henan International Joint Laboratory of MXene Materials Microstructure, College of Physics and Electronic Engineering, Nanyang Normal University, Nanyang 473061, China\\
$^3$ School of Mechanical Engineering and Electronic Information, China University of Geosciences, Wuhan 430074, People's Republic of China\\
$^4$ Research Center on Vector Optical Fields, Institute of Optics and Electronics, Chinese Academy of Sciences, Chengdu 610209, People's Republic of China}

\date{\today}

\begin{abstract}
{The light scalar field with a coupling to standard model particles provide a possible source of the long-range Yukawa forces or violation of the weak equivalence principle, which can be potentially explored by precision gravity experiments.} We describe the searches for {such light scalar fields} with the three types of gravity experiments, including the $G$-measurement experiments, Inverse-Square Law (ISL) experiments, and equivalence principle experiments. We investigate the potential influences of the {scalar field} as a function of its mass, and focus on the experimental constraints from torsion-balance gravity experiments.
HUST-18 $G$-measurement torsion-balance experiments place bounds on {the photon coupling and electron coupling} at up to $\Lambda_{\gamma}=7\times10^{17}$ GeV and $\Lambda_{e}=1\times10^{17}$ GeV in the mass ranges $10^{-9}-10^{-4}$ eV.
Results from the ISL experiments by the Universities of Washington, Stanford, IUPUI, HUST, Colorado, Irvine, Yale and others allow us to set limits on the photon coupling and electron coupling at up to $\Lambda_{\gamma}=5\times10^{17}$ GeV and $\Lambda_{e}=3\times10^{16}$ GeV for {scalar field mass ranges} between $10^{-5}$ and $10^{-1}$ eV. Additionally, we also discuss the limits from equivalence principle experiments, and $MICROSCOPE$ final result updates the constrains on the {coupling parameters} at up to $\Lambda_{\gamma}=7\times10^{22}$ GeV and $\Lambda_{e}=4\times10^{21}$ GeV for mass ranges $\lesssim 10^{-13}$ eV. These results contribute experimental constraints to relatively unexplored mass regions of {light scalar field} parameter space and improve upon previous limits in some mass ranges. This work paves the way for {long-range Yukawa forces mediated by light scalar fields} in future high-precision gravity experiments.
$G$-value measurement, Inverse-Square law tests
\end{abstract}

\maketitle
\section{introduction}
{The nature of the dark matter (DM) in the Universe is mysterious,} although the astrophysical and cosmological observations have strong indirect evidences of DM existence \cite{clowe2006direct,ade2016planck}. Many DM candidates and laboratory experiments have been proposed to shed light on the nature of dark matter \cite{bertone2010particle,armengaud2012search,baum2023mineral,PhysRevD.88.012002,PhysRevD.104.115033,buettner2022neutrinos,djuvsland2023inverse,ren2015probing}. The class of DM candidates that has attracted the most attention is weakly interacting massive particles (WIMPs) \cite{roszkowski2018wimp,hooper2007dark,oikonomou2007direct,PhysRevD.75.023513}. Another very popular and novel class of DM candidates is axion and axion-like particles \cite{PhysRevLett.124.251802,hall2020axion,PhysRevD.103.044036,PhysRevD.106.044041,PhysRevD.107.064071,marsh2016axion}, which motivate new ways for DM searches. To make progress on the dark matter problem, current efforts emphasize experimental diversification, probing different models over a wide mass range \cite{bertone2018new,RevModPhys.90.025008}.
The bosonic DM scenario is one of the most popular candidates, which can have a sub-eV mass and behave more like a classical wave than a particle. Among the motivated candidates in this category, QCD axion and axion-like particles mainly have derivative interaction with Standard Model (SM) particles, which are parity-odd pseudo-Nambu-Goldstone bosons (PNGBs) \cite{fan2016ultralight,abe2021direct,abe2020pseudo,PhysRevD.104.035011,ge2022x,addazi2022qcd,PhysRevLett.119.191801}. For light, parity-even bosonic candidates, dilatons are particularly prominent \cite{taylor1988dilaton,damour1994string,PhysRevD.91.015015}.
In those models, a light scalar field is introduced and coupled to standard-model fields. Such scalar fields are ubiquitous in theories with high dimensions, such as the string theory \cite{PhysRevLett.89.081601,PhysRevD.65.023508}. {In addition, the massive scalar fields also are introduced in the scalar-tensor theories.}

{The existence of a light scalar field implies some new physics beyond the general relativity or standard model.}
At the low-energy region, coherent oscillations of fundamental constants of nature, {long-range Yukawa forces} and violations of the equivalence principle (EP) are caused by dilaton scalar fields. This has motivated a multitude of experimental searches in a wide mass range. These searches include experiments using the atomic clocks in the mass range of $10^{-24}-10^{-14}$ eV \cite{PhysRevLett.115.011802,PhysRevLett.117.061301,PhysRevLett.125.201302,wcislo2018new,PhysRevLett.129.241301}, atom multigradiometry in the mass range of $10^{-17}-10^{-14}$ eV \cite{PhysRevD.107.055002}, the comparisons by various oscillators in the mass range of $10^{-19}-10^{-13}$ eV \cite{PhysRevLett.126.071301}, EP violation tests like $MICROSCOPE$ mission in the mass range of $10^{-18}-10^{-11}$ eV \cite{PhysRevLett.120.141101}, as well as experiments utilizing magnetometer, optical cavity, interferometer and atomic spectroscopy in the mass range of $10^{-14}-10^{-9}$ eV \cite{PhysRevLett.126.051301,PhysRevD.107.075033,PhysRevLett.123.141102}. {To explore the parameter space of scalar fields more extensively}, other experiments and proposals have been considered, including the gravitational-wave detectors \cite{PhysRevD.100.123512,PhysRevLett.118.021302,PhysRevD.97.075020,PhysRevResearch.1.033187,vermeulen2021direct,PhysRevD.107.043004}, resonant-mass detectors \cite{PhysRevLett.116.031102}, alkali-atoms hyperfine transitions, capacitors \cite{PhysRevD.107.015008,PhysRevD.106.055037,PhysRevD.105.083533,PhysRevD.103.083535,PhysRevD.98.102002}, cosmological observations \cite{zhang2022imprints,addazi2018testing,bird2023snowmass2021,khlopov1985gravitational} etc.
For larger masses, laboratory gravity experiments, such as tests of the universality of free fall or Inverse-Square Law (ISL) tests, become relevant. These gravity experiments are highly sensitive to the tests of fundamental physics, such as Lorentz symmetry \cite{PhysRevD.74.045001,PhysRevLett.117.071102,PhysRevLett.122.011102}, chameleon model \cite{PhysRevD.103.104005}, and equivalence principle \cite{PhysRevD.61.022001}. For example, the tests of gravitational ISL have been performed by the University of Washington \cite{PhysRevLett.86.1418,PhysRevLett.98.021101,PhysRevD.70.042004,PhysRevLett.124.101101}, Stanford \cite{PhysRevLett.90.151101,PhysRevD.78.022002}, IUPUI \cite{PhysRevLett.116.221102}, HUST \cite{PhysRevLett.98.201101,PhysRevLett.108.081101,PhysRevLett.116.131101}, Colorado \cite{long1999experimental,long2003upper}, Irvine \cite{PhysRevD.32.3084}, Yale \cite{PhysRevLett.107.171101} among others, whose results can be used to place the constraints on the {scalar field couplings}. In addition, we find that utilizing two or multiple $G$-measurement gravity experiments can be proposed to constrain {the scalar field couplings}. The addition of $G$-measurement gravity experiments can contribute to the diversity of experimental limits on {scalar field coupling}. In this paper, we present {the searches for long-range Yukawa forces mediated by light scalar fields} with three types of gravity experiments including the $G$-measurement experiments, Newton ISL experiments, and equivalence principle experiments, which can provide the constraints on scalar field over a broad mass range.

The rest of this paper is organized as follows. In Sec.\ref{sec2}, we review the theory of dilaton scalar field and the theory leads to a scalar Yukawa interaction. Then, {the influences of scalar fields} on the $G$-measurement, ISL test, and equivalence principle test gravity experiments are discussed in sub-Sec.\ref{sec23}, \ref{sec21} and \ref{sec22}, respectively. In Sec.\ref{sec3}, we present the experimental constraints on {scalar coupling parameter spaces} by reanalyzing the results of the HUST-18 $G$-measurement experiment, HUST-20 ISL test experiments, and $MICROSCOPE$ equivalence principle experiment.
The conclusion is given in Sec.\ref{sec4}. The Appendix \ref{appe3} presents some calculation details for ISL test. The methods of calculating the {coupling} parameter in the HUST-18 and HUST-20 experiments are given in Appendix \ref{appe1} and Appendix \ref{appe2}, respectively.

\section{{A light scalar field} and gravity effects}\label{sec2}

The theory and general phenomenology of {dilaton-like scalar field} have been studied in, e.g., Refs. \cite{PhysRevD.82.084033,PhysRevD.91.015015}. In this section, we focus on studying the gravity effects of {such a scalar field} coupled to the SM particles. We show that $G$-value measurement, ISL test, and equivalence principle test experiments can be used to place the bounds on the coupling parameters of scalar field. Considering the couplings between {a scalar field $\phi$ and the SM particles and fields, the $\phi$-dependent action is parameterized as follows \cite{PhysRevD.82.084033,PhysRevD.91.015015}
\begin{eqnarray}\label{dm1}
  \mathcal{L}_{\phi}=\kappa \phi \Big{[}  \frac{d_{e}}{4e^{2}}F_{\mu\nu}F^{\mu\nu}-\frac{d_{g}\beta_{3}}{2g_{3}}F^{A}_{\mu\nu}F^{A\mu\nu}
  -\sum_{i=e,u,d}(d_{m_{i}}+\gamma_{m_{i}}d_{g})m_{i}\bar{\Psi}_{i}\Psi_{i}\Big{]},
\end{eqnarray}
where $\kappa={\sqrt{4\pi}}/{M_{\text{Pl}}}$, $g_{3}$ is the QCD gauge coupling, $\beta_{3}$ is the QCD beta function, $F_{\mu\nu}$ is the electromagnetic Faraday tensor, $F^{A}_{\mu\nu}$ is the gluon strength tensor, $\gamma_{m_{i}}$ are the anomalous dimensions of the electron, $u$ quark and $d$ quark, $\Psi_{i}$ denotes the fermion spinors, and $d_{a}$ ($a=g, e, m_{u}, m_{d}, m_{e}$) are the dimensionless coupling coefficients that describe the coupling strength of scalar field to the SM particles. Note that these coupling coefficients sometimes are defined by another convention with dimensional coefficients $\Lambda_{a}$, such as in the atomic clock experiments and interferometer-like experiments, and the relation between two set coefficients is given by $\Lambda_{a}=M_{\text{Pl}}/(\sqrt{4\pi}d_{a})$.

For the relevant low-energy couplings, the coupling constants of the scalar field in the Lagrangian density will cause a $\phi$ dependence in the fundamental constants of nature, as follows
\begin{eqnarray}
% \nonumber % Remove numbering (before each equation)
  \Lambda_{3}(\phi) &=& \Lambda_{3}(1+d_{g}\kappa\phi) \nonumber\\
  \alpha(\phi) &=& \alpha(1+d_{e}\kappa\phi) \nonumber\\
  m_{i}(\phi) &=& m_{i}(1+d_{m_{i}}\kappa\phi),
\end{eqnarray}
where $\Lambda_{3}$ is the QCD confinement scale and $\alpha$ is the fine structure constant.
Considering the variations of the fine structure constant and masses of proton and electron, it can lead to changes in the frequency of atomic clocks since the quantum transition frequencies are dependent on the fine structure constant and proton-electron mass ratio. This motivates {many experimental searches involving various atomic clocks}.
Considering the variations of masses of fundamental particles and fine structure constant, it can give rise to the violations of equivalence principle since the body's acceleration is dependent on the body composition in the presence of scalar field. In addition, the scalar field may show up in current tests of the gravitational inverse-square law at sub-millimeter ranges. These facilitate the gravity experimental searches, e.g., the universality of free fall tests and the ISL test experiments \cite{PhysRevD.82.084033,damour1992tensor,wagner2012torsion,PhysRevLett.120.141101,PhysRevD.98.064051}.

Considering an atom, its mass can be decomposed as $m(Z,N)=Zm_{p}+Nm_{n}+Zm_{e}+E^{\text{binding}}$, where $Z$ is the atomic number, $N$ is the number of neutrons and $E^{\text{binding}}$ consists of strong interaction and electromagnetic binding energies.
From the scalar field couplings to the SM fields in the action (\ref{dm1}), one deduces the variations in the fine structure constant, the masses of quarks and electrons, and atomic binding energies, then obtains the total changes of atom mass $m(Z,N)$. In order to isolate EP violation, it is conventional to introduce the symmetric combination $\hat{m}=(m_{d}+m_{u})/2$ and antisymmetric combination $\delta m=m_{d}-m_{u}$ of the quark masses, which depend on scalar field $\phi$ as following form
\begin{eqnarray}
% \nonumber % Remove numbering (before each equation)
  \hat{m}(\phi) &=& \hat{m}(1+d_{\hat{m}}\kappa\phi),\nonumber \\
  \delta m(\phi) &=& \delta m(1+d_{\delta m}\kappa\phi),
\end{eqnarray}
where their dilaton-coupling coefficients are given by
\begin{eqnarray}
% \nonumber % Remove numbering (before each equation)
  d_{\hat{m}} &=& \frac{d_{m_{d}}m_{d}+d_{m_{u}}m_{u}}{m_{d}+m_{u}} \nonumber \\
  d_{\delta m} &=& \frac{d_{m_{d}}m_{d}-d_{m_{u}}m_{u}}{m_{d}-m_{u}},
\end{eqnarray}
respectively. As a result of action (\ref{dm1}), the existence of a scalar field $\phi$ modifies the gravity interaction between a point mass $m_{A}$ and a point mass $m_{B}$, which is given in the form of Yukawalike coupling \cite{PhysRevD.91.015015,PhysRevD.82.084033}
\begin{equation}\label{dm3}
  V \left( r\right) =-\frac{Gm_{A}m_{B}}{r_{AB}}\left( 1+\alpha_{A}\alpha_{B}e^{-m_{\phi}r_{AB}}\right),
\end{equation}
where $m_{\phi}$ is the mass of the scalar field $\phi$, $G$ is the Newton gravitational constant, $r_{AB}$ is the distance of point masses $m_A$ and $m_B$, and the unit system is natural unit. In Eq.(\ref{dm3}), {we assume that the Yukawa interaction is caused only by the light scalar field. We focus on the long-range forces and gravity effects mediated the light scalar field.} The parameters $\alpha_{A}$ and $\alpha_{B}$ represent scalar field coupling strengths to masses $A$ and $B$, which are dependent on the components of point masses $m_A$ and $m_B$.
Here, to avoid confusion with the fine structure constant, the coupling strengths in the potentials are denoted by the variable $alpha$ with the corresponding subscript such as $\alpha_{b}$, and the fine structure constant is represented by the $\alpha$ without subscript.
The dimensionless coupling parameter $\alpha_{b}$ can be expressed as
\begin{equation}\label{dm4}
  \alpha_{b}=\frac{\partial \ln [\kappa m_{b}(\kappa\phi)]}{\partial \kappa\phi}=d_{g}+\bar{\alpha}_{b},
\end{equation}
where the parameter $\bar{\alpha}_{b}$ is given by
\begin{eqnarray}\label{dm5}
  \bar{\alpha}_{b}=\Big{[}(d_{\hat{m}}-d_{g})Q_{\hat{m}}+(d_{\delta m}-d_{g})Q_{\delta m}
  +(d_{m_{e}}-d_{g})Q_{m_{e}} +d_{e}Q_{e}\Big{]}_{b},
\end{eqnarray}
where the dilaton charges are $Q_{e}=\frac{\partial \ln m_{b}}{\partial \ln \alpha}$, $Q_{m_{e}}=\frac{\partial \ln m_{b}}{\partial \ln {m}_{e}}$, $Q_{\hat{m}}=\frac{\partial \ln m_{b}}{\partial \ln \hat{m}}$, and $Q_{\delta m}=\frac{\partial \ln m_{b}}{\partial \ln \delta m}$ that are given by
\begin{equation}\label{dilaq1}
  Q_{\hat{m}}=\left( 9.3-\frac{3.6}{A^{1/3}}-2\frac{(A-2Z)^{2}}{A^{2}}-1.4\times10^{-2}\frac{Z(Z-1)}{A^{4/3}}\right)F_{A}\times10^{-2},
\end{equation}

\begin{equation}\label{dilaq2}
  Q_{\delta m}=\left( 1.7 \frac{A-2Z}{A}\right)F_{A}\times10^{-3},
\end{equation}

\begin{equation}\label{dilaq3}
  Q_{m_{e}}=\left( 5.5 \frac{Z}{A}\right)F_{A}\times10^{-4},\nonumber
\end{equation}

\begin{equation}\label{dilaq4}
  Q_{e}=\left( -1.4+8.2  \frac{Z}{A}+7.7\frac{Z(Z-1)}{A^{4/3}} \right)F_{A}\times10^{-4},
\end{equation}
where $F_{A}=A m_{u}/m_{amu}=1+O(10^{-4})$, $A$ is the mass number, $Z$ is the atomic number, and $m_{amu}=931$ MeV is the atomic mass unit, $m_{b}$ is the total mass of an atom.
In the Newton ISL test experiments, one can detect the long-range Yukawa forces by looking for a departure from $r^{-1}$ in potential.

Considering the experiments of EP violation, it is convenient to decompose parameter $\alpha_{b}$ into a composition-independent part $d_{g}^{*}$ and a composition-dependent part $\bar{\alpha}_{b}^{*}$
\begin{equation}\label{dmlamb}
  \alpha_{b}=d_{g}^{*}+\bar{\alpha}_{b}^{*}=d_{g}^{*}+\Big{[}(d_{\hat{m}}-d_{g})Q^{*}_{\hat{m}}+(d_{\delta m}-d_{g})Q^{*}_{\delta m}
  +(d_{m_{e}}-d_{g})Q^{*}_{m_{e}} +d_{e}Q^{*}_{e}\Big{]}_{b}.
\end{equation}
This composition is based on the fact that for most elements, the mass number and atomic number satisfy relation $A\sim2Z$. Then, the composition-independent part $d_{g}^{*}$ is given by
\begin{equation}\label{dmlamb1}
  d_{g}^{*}=d_{g}+0.093(d_{\hat{m}}-d_{g})+[2.75(d_{m_{e}}-d_{g})+2.7d_{e}]\times10^{-4},
\end{equation}
and the composition-dependent part $\bar{\alpha}_{b}^{*}$ is characterized by new dilaton charges
\begin{equation}\label{dilaqb1}
  Q^{*}_{\hat{m}}=\left(-\frac{3.6}{A^{1/3}}-\frac{2(A-2Z)^{2}}{A^{2}}-0.014\frac{Z(Z-1)}{A^{4/3}}\right)\times10^{-2},\nonumber
\end{equation}

\begin{equation}\label{dilaqb2}
  Q^{*}_{\delta m}= 1.7\times10^{-3} \frac{A-2Z}{A},\nonumber
\end{equation}

\begin{equation}\label{dilaqb3}
  Q^{*}_{m_{e}}= -2.75\times10^{-4} \frac{A-2Z}{A},\nonumber
\end{equation}

\begin{equation}\label{dilaqb4}
  Q^{*}_{e}=\left(-4.1 \frac{A-2Z}{A}+7.7\frac{Z(Z-1)}{A^{4/3}}\right)\times10^{-4}.
\end{equation}
Clearly, the values of dilaton charges depend on only the composition of the considered object and these values are non-zero for most bodies. Therefore, it leads to a violation of the equivalence principle. By combining materials and dilaton charges of test masses, the existing experiments of testing equivalence principle can be used to place bounds on the couplings of the scalar field $\phi$.

\subsection{$G$-measurement gravity experiments}\label{sec23}

In this subsection, we propose {a search for long-range Yukawa forces mediated by light scalar fields} with the $G$-measurement gravity experiments. As the fundamental physical constant, $G$ value has been measured by numerous gravity experiments, which are potential methods to explore {the possible existence of the light scalar fields}. From the Yukawalike coupling (\ref{dm3}), the {scalar-field effect} can be absorbed by the effective gravitational constant $G_{\text{eff}}$
\begin{equation}\label{gmdk1}
  G_{\text{eff}}=G_{0}\left( 1+\alpha_{A}\alpha_{B}e^{-m_{\phi}r_{AB}}\right),
\end{equation}
where $G_{0}$ is the Newton gravitational constant. Obviously, in the presence of {a scalar field}, the measured value of the $G$ will depend on the experimental materials and setup, leading to different $G$-measurement values in different experiments. However one measured value of the $G$ can't be used to detect {the scalar-field effects}. The difference of $G$-value measurements performed by different experiments may reflect the influence of {the scalar fields}. By utilizing various $G$-measurement experiments, it is possible to {set constraint on the coupling parameters space}.
Considering that two experiments 1 and 2 obtained the $G$-measurement values $G_{1}$ and $G_{2}$, the measurement values can be split into the Newton gravitational constant part $G_{0}$ and a corrected part induced by {the scalar fields}. From two $G$ values, we can structure a ratio
\begin{equation}\label{dmg1}
\left| \frac{G_{1}}{G_{2}}-1 \right| = \Delta {G_{12}}=\left| \frac{G_{0}(1+\mathcal{M}_{1}(m_{\phi},d_{i}))}{G_{0}(1+\mathcal{M}_{2}(m_{\phi},d_{i}))}-1 \right|,
\end{equation}
where $\Delta_{G_{12}}$ is calculated parameter from the experimental measured values and corresponding uncertainties; $\mathcal{M}_{1}(m_{\phi},d_{i})$ and $\mathcal{M}_{2}(m_{\phi},d_{i})$ represent the {scalar-field} influences in the $G$-measurement experiments 1 and 2, respectively, which depend on the experimental materials and apparatus. Different $G$-measurement gravity experiments have different values of parameter $\mathcal{M}(m_{\phi},d_{i})$. One can search for possible {scalar-field} effects by using two $\mathcal{M}(m_{\phi},d_{i})$ and $G$-value measurements. Therefore, combining $G$-measurement values from multiple experiments can {set constraint on the parameter spaces}. Several $G$-value measurements with extremely high precision were reported by the various gravity experiments\cite{tiesinga2021codata}, in which {scalar-field} effects were dependent on the corresponding experiment apparatus. These $G$-measurement gravity experiments can diversify {experimental searches for long-range Yukawa forces mediated by light scalar fields}. Based on two specific experiments, {a experimental constraint} can be presented by using Eq.(\ref{dmg1}).

\subsection{Inverse-Square Law test gravity experiments}\label{sec21}

In this subsection, we focus on the {long-range Yukawa effects induced by light scalar fields with} the Inverse-Square Law test gravity experiments. For the small mass of {scalar field}, the Solar system planets motions and satellites orbits can be used to place on the bounds of scalar field couplings. This scenario can be treated as a test body orbiting the massive central body with the Kepler orbit $r=a(1-e^{2})/(1+e\cos{f})$, where $a$ is the semimajor axis, $e$ is the eccentricity, $f$ is the true anomaly. The force induced by scalar field coupling can be considered a small perturbation to the Newtonian gravity. Then, the {perturbing force} leads to changes in the Kepler orbital parameters.
From the Yukawa coupling (\ref{dm3}), the acceleration is given by
\begin{equation}\label{fif1}
  \textbf{\emph{a}}_{\text{sf}}=-\frac{GM_{S}}{r^{3}}\alpha_{S}\alpha_{T}I\left(m_{\phi}r_{S}\right)\left(1+m_{\phi}r\right)e^{-m_{\phi}r}\textbf{\emph{r}}
\end{equation}
with function
\begin{equation}\label{fif01}
  I(x)=3\frac{x\cosh{x}-\sinh{x}}{x^{3}},
\end{equation}
where $\alpha_{T}$ is the coupling parameter of test body, $\alpha_{S}$ is the coupling parameter of source body, $M_{S}$ is the mass of source body, $r_{S}$ is the radius of the massive central body and $\textbf{\emph{r}}$ is the position vector pointing from the massive body to the orbiting test body.
From the disturbing acceleration (\ref{fif1}), the {Yukawa} force has only the radial component $\mathcal{A}$ affecting in-plane orbital motions. Thus, the orbital inclination $i$ and longitude of the ascending node $\Omega$ are not affected by {perturbing force} (see Appendix.{\ref{appe3}}). The secular-variation Kepler parameters concentrates on the argument of perigee $\omega$, which is given by (see the Appendix.{\ref{appe3}})
\begin{equation}\label{fif001}
  \left\langle\frac{d \omega}{dt}\right\rangle=\left\langle-\frac{\sqrt{1-e^{2}}}{nae}\cos{f}\mathcal{A}  \right\rangle ,
\end{equation}
where $n=\sqrt{GM_{S}/a^3}$ is Keplerian mean motion.

In the case of $1/m_{\phi}\gg a$, the exponential term in the acceleration (\ref{fif1}) can be approximated as $e^{-m_{\phi}r}\simeq1-m_{\phi}r$. Inserting this approximation into the Gauss equations, one can obtain the long-term changes after the averaging over an orbital period. For the order of $e^{2}$, the secular variation of the argument of perigee $\omega$ is given by
 \begin{equation}\label{fif8}
  \left\langle\frac{d \omega}{dt}\right\rangle=\alpha_{S}\alpha_{T}\frac{n}{(1-e^2)^{3/2}}I(m_{\phi }r_{S}).
\end{equation}

In the case of $1/m_{\phi}\sim a$, we assume the relation $1/m_{\phi}\gtrsim ae$ since the small eccentricity $e$. Considering the Kepler orbit equation $r=a(1-e\cos{E})$, the exponential term in the acceleration (\ref{fif1}) can be approximated as
\begin{equation}\label{fif09}
  e^{-m_{\phi}r}\simeq e^{-m_{\phi}a}\left(1+m_{\phi}ae\cos{E}\right)
\end{equation}
where $E$ is the eccentric anomaly of orbit. Inserting this approximation into the Gauss equations and averaging them with respect to $E$, one can obtain the long-term changes for the Kepler parameters. For the order of $e^{2}$, the secular variation of the argument of perigee $\omega$ is given by \cite{iorio2007constraints}
\begin{equation}\label{fif9}
  \left\langle\frac{d \omega}{dt}\right\rangle=\alpha_{S}\alpha_{T}\frac{n(m_{\phi}a)^{2}}{2}I(m_{\phi }r_{S})e^{-m_{\phi}a}
\end{equation}

In the case of $1/m_{\phi}\ll a$, the exponential term in the acceleration (\ref{fif1}) tends to 0 and orbit motions are not sensitive to the {long-range forces induced by the scalar fields}. The corresponding mass ranges can be neglected. For the ISL test experiments, the formulas of Eqs.(\ref{fif8}) and (\ref{fif9}) are useful. They can be straightforwardly used to constrain the parameter spaces of {light scalar fields} from the orbital measurements of the satellites, planets and other celestial bodies. For example, considering the mass ranges $m_{\phi}\ll 10^{-18}$ eV, the precession of the longitude of the perihelion of Saturn $\Delta \varpi =-0.006\pm0.002$ arcsec/century can be considered as the bound for coupling parameter $\alpha_{\odot}\alpha_{\text{Sat}}\leq (13.8\pm4.6)\times10^{-10}$, where $\alpha_{\odot}$ is the coupling parameter of the Sun and $\alpha_{\text{Sat}}$ is the coupling parameter of the Saturn.

In addition, the observational accuracy of the perihelion precession of the planets is $10^{-4}-10^{-3}$ arcsec/century. For a pair of planets in the Solar System, we can use Eqs.(\ref{fif8}) and (\ref{fif9}) to give the equations
\begin{eqnarray}\label{dmcc1}
% \nonumber % Remove numbering (before each equation)
 {\Delta \varpi_{A} }-{\Delta \varpi_{B}}&=& \alpha_{\bigodot}I(m_{\phi }r_{\odot})\left(\frac{\alpha_{A}n_{A}}{(1-e_{A}^2)^{3/2}}-\frac{\alpha_{B}n_{B}}{(1-e_{B}^2)^{3/2}} \right) , \text{for}\quad 1/m_{\phi}\gg a,\nonumber \\
  {\Delta \varpi_{A} }-{\Delta \varpi_{B}} &=& \alpha_{\bigodot}I(m_{\phi }r_{\odot})\left(\frac{\alpha_{A}n_{A}(m_{\phi}a_{A})^{2}}{2}e^{-m_{\phi}a_{A}}
  -\frac{\alpha_{B}n_{B}(m_{\phi}a_{B})^{2}}{2}e^{-m_{\phi}a_{B}}\right), \text{for} \quad 1/m_{\phi} \sim a.
\end{eqnarray}
If the components of two planets are significantly different, such as Earth and Jupiter, Eq.(\ref{dmcc1}) can be used to set limits on the {parameter space of the scalar field}.

Another type of ISL test gravity experiments is the test of the gravitational inverse-square law in the laboratories. In this type of gravity experiments, the torsion balance is a very powerful tool to test the gravitational ISL at short ranges. By using the schemes of the separation modulation or rotation modulation, the Yukawa effects to be measured are kept in the most sensitive direction of the pendulum. After compensating the Newtonian torque, the experiments can obtain the limit of the Yukawa interaction from the residual torque. In the presence of Yukawa force induced by {the scalar fields}, the motion equation of the torsion pendulum is characterized by
\begin{equation}\label{fif11}
   I\ddot{\theta}+k\theta=\tau+\tau_{\text{sf}},
\end{equation}
where $I$ is the inertial moment of the pendulum, $\theta$ is the twist angle of the pendulum, $k$ is the spring constant, $\tau$ represents torque in the Newtonian gravity frame, and $\tau_{\text{sf}}$ represents the torque signal of {Yukawa interaction} that is given by the numerical integration
\begin{equation}\label{dm6dm}
  \tau_{\text{sf}}(m_{\phi})=\frac{\partial}{\partial\theta}\int_{V_{i},V_{j}}
  \frac{G\rho_{i}\rho_{j}}{r}\alpha_{i}\alpha_{j}e^{-m_{\phi}r}d^{3}r_{i}d^{3}r_{j},
\end{equation}
where $\rho_{i}$ and $\rho_{j}$ are, respectively, the densities of the components on the pendulum $i$ and attractor$j$, $r$ is the distance between the volume element of the bodies $i$ and $j$, $\alpha_{i}$ and $\alpha_{j}$ are the coupling parameters of the bodies $i$ and $j$, respectively. For the experimental pendulum and attractor, the coupling parameters $\alpha_{i}$ and $\alpha_{j}$ can be treated as the constant given by Eq.(\ref{dm4}). By using the compensation masses to compensate the Newtonian torque $\tau$, the measurement torque is sensitive to the torque induced by {the scalar fields}, which can be used to place the bounds on the {coupling parameter spaces}. The torsion balance experiments with E\"{o}t-Wash and HUST groups are highly sensitive to the parameter spaces in the mass ranges of $10^{-7}-10^{-2}$ eV.

\subsection{Equivalence principle gravity experiments}\label{sec22}

In this subsection, we consider {a search for long-range Yukawa forces mediated by light scalar fields with} the equivalence principle experiments. The most precision tests of equivalence principle are to compare the accelerations of two test bodies of different composition or internal structure in an external gravitational field.
The violation of equivalence principle is described by the E\"{o}tv\"{o}s parameter, defined by
\begin{equation}\label{epdk1}
  \eta=2\frac{|\textbf{\emph{a}}_{A}-\textbf{\emph{a}}_{B}|}{|\textbf{\emph{a}}_{A}+\textbf{\emph{a}}_{B}|},
\end{equation}
where $\textbf{\emph{a}}_{A}$ and $\textbf{\emph{a}}_{B}$ are the accelerations of the test bodies $A$ and $B$ relative to the central attractor, respectively. As the existence of {a light scalar field} can induce the violation of the equivalence principle, Eq.(\ref{epdk1}) can be directly translated into the constraints on {coupling parameter spaces}.

Most tests of equivalence principle experiments are based on the external gravitational field of the Earth or the Sun. In this scenario, we assume that the external central body $S$ is a spherically symmetric body with the radius $R_{S}$. Considering the Yukawalike coupling (\ref{dm3}), the acceleration of the test body $A$ is given by
\begin{equation}\label{epdk2}
  \textbf{\emph{a}}_{A}=\textbf{\emph{g}}_{S}-\alpha_{S}\alpha_{A}\int_{S}\frac{G\rho_{S}}{r^{3}}\left(1+m_{\phi}r\right)e^{-m_{\phi}r}\textbf{\emph{r}} d^{3}r_{S},
\end{equation}
where the first term is the Newtonian gravity acceleration of source body $S$, and the second term is the acceleration relevant to the material of test body and source body.

If the external source body is a uniform-density spherically body, the {scalar-field} part of acceleration becomes
\begin{equation}\label{epdk3}
  \textbf{\emph{a}}_{A,\text{sf}}=\alpha_{S}\alpha_{A}I^{(1)}\left(m_{\phi}R_{S}\right)\left(1+m_{\phi}r\right)e^{-m_{\phi}r}\textbf{\emph{g}},
\end{equation}
where the function $I^{(1)}\left(m_{\phi}R_{S}\right)=I\left(m_{\phi}R_{S}\right)$ represents uniform density for source body $S$. In the case where the distance from the test body to the source body is much larger than the size $R_{S}$ of the source body, it is accurate enough to use Eq.(\ref{epdk3}) for limiting the {coupling parameter spaces}, such as the test of weak equivalence principle with Lunar Laser Ranging.

The assumption of uniform density for a source body $S$ is too ideal. Especially for Earth, it's interior contains complex structures including the core, mantle, crust, etc. The density of the core, mantle, and crust is different. So the Earth should be considered as a $n$-layers spherical body. We consider a model assuming that the source body $S$ is a spherically symmetric body composed of $n$ layers, and each layer has its corresponding density. The gravity acceleration is given by integrating over $n$ layers of body $S$. In this $n$-layers model, the {scalar-field} acceleration of a test mass $A$ is given by
\begin{equation}\label{epdk4}
  \textbf{\emph{a}}_{A,\text{sf}}=\alpha_{S}\alpha_{A}I^{(n)}\left(m_{\phi}R_{S}\right)\left(1+m_{\phi}r\right)e^{-m_{\phi}r}\textbf{\emph{g}},
\end{equation}
with the function
\begin{equation}\label{epdk5}
  I^{(n)}\left(m_{\phi}R_{S}\right)=\sum_{i}^{n}\frac{M_{i}}{M_{S}}\frac{R_{i}^{3}I(m_{\phi}R_{i})-R_{i-1}^{3}I(m_{\phi}R_{i-1})}{R^{3}_{i}-R^{3}_{i-1}}
\end{equation}
where $M_{S}$ is the total mass of the source body $S$, $M_{i}=4\pi\rho_{i}(R^{3}_{i}-R^{3}_{i-1})/3$ is the mass of $n$th layer with density $\rho_{i}$, $R_{i}$ is the radius of the $n$th layer, $R_{0}=0$ and $R_{n}=R_{S}$. For the EP test experiments in the laboratory, Eq.(\ref{epdk4}) is more accurate than Eq.(\ref{epdk3}) because the Earth is not a uniform-density body. To demonstrate the effect of $n$ layers, we assume that the apparatus performs the EP test above the Earth's surface with height $h=1$ m. We evaluated the effective corrected factor $ I^{(n)}_{\text{eff}}= I^{(n)}\left(R_{E}/\lambda_{\phi}\right)\left(1+m_{\phi}r\right)e^{-(R_{E}+h)/\lambda_{\phi}}$ for the Earth model with one (uniform density), two and eleven layers, as shown in Fig.\ref{fig:ep1}. The uniform-density Earth is represented by the black line and the 11-layer Earth is represented by the red line. For the ranges $\lambda_{\phi}>10^{6}$ m, different Earth models are consistent. For the ranges $\lambda_{\phi} < 10^{6}$ m, $n$-layers model is more accurate that is more suitable. For larger mass ranges, the $n$-layers effects are more significant.

\begin{figure}
\includegraphics[width=0.5\textwidth]{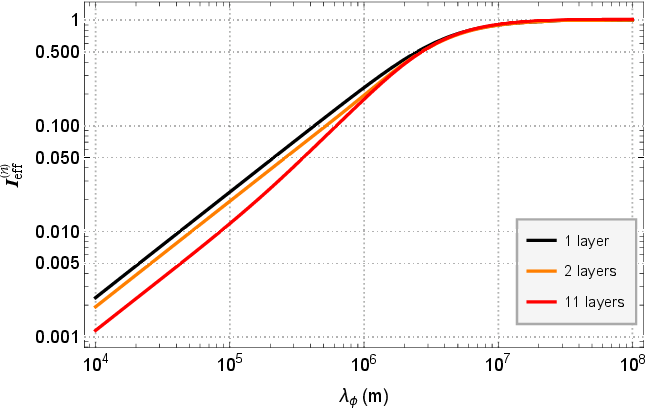}
\caption{\label{fig:ep1} The evaluation of the effective correction factor $I^{(n)}_{\text{eff}}$ for the Earth model. The black, orange, and red curves represent the Earth model with 1 (uniform density), 2, and 11 layers. }
\end{figure}

In the EP test experiments, the composition-independent part of the source body dominates over the composition-dependent part, and the composition-dependent part of test bodies $A$ and $B$ dominate the violating effects. From Eqs.(\ref{dmlamb})-(\ref{dilaqb4}) and (\ref{epdk4}), the E\"{o}tv\"{o}s parameter defined by Eq.(\ref{epdk1}) becomes
\begin{equation}\label{epdk6}
  \eta=\Delta_{A,B}\alpha_{S} I^{(n)}({m_{\phi}}R_{S})(1+m_{\phi }r) e^{-m_{\phi }r}=\Delta_{A,B}\alpha_{S} I^{(n)}_{\text{eff}}
\end{equation}
with
\begin{equation}\label{epdk7}
  \Delta_{A,B}\alpha_{S}=(\alpha_{A}-\alpha_{B})\alpha_{S} \approx \Delta Q^{*}_{\hat{m}}D_{\hat{m}}+\Delta Q^{*}_{\delta m}D_{\delta m}
  +\Delta Q^{*}_{m_{e}}D_{m_{e}} +\Delta Q^{*}_{e}D_{e}
\end{equation}
where the parameters are defined as $D_{\hat{m}}\equiv d^{*}_{g}(d_{\hat{m}}-d_{g})$, $D_{\delta m}\equiv d^{*}_{g}(d_{\delta m}-d_{g})$, $D_{m_{e}}\equiv d^{*}_{g}(d_{m_{e}}-d_{g})$, and $D_{e}\equiv d^*_{g}d_{e}$. In the typical experiments, it satisfies that $ \Delta Q^{*}_{\hat{m}}\gg \Delta Q^{*}_{\delta m}$ and $\Delta Q^{*}_{e}\gg \Delta Q^{*}_{m_{e}}$ so that the existing experimental bounds on $\eta$ can be used to constrain the parameter space ($D_{\hat{m}}, D_{e}, m_{\phi}$) for a given mass of {a scalar field}. Moreover, assuming only one nonvanishing coupling at a time, the experimental bound on $\eta$ also can be translated into the limitation on individual coupling parameter $d_{a}$
\begin{equation}\label{epdk8}
  d_{a}=\sqrt{\frac{\eta}{\Delta Q_{a} Q_{a,S}I^{(n)}_{\text{eff}}({m_{\phi}}R_{S})}},
\end{equation}
where $Q_{a,S}$ is the dilaton charge of the source body.

\section{The parameter constraints on the scalar field {coupling parameters}}\label{sec3}

There are several coupling coefficients to be limited. Due to the combination of coefficients in measurements, it is difficult to set the limits on the individual coupling coefficients. As the analyses in most papers of atomic sensors \cite{PhysRevLett.115.011802,PhysRevLett.117.061301,PhysRevLett.114.161301,PhysRevLett.115.201301,PhysRevLett.116.031102} and \emph{MICROSCOPE} experiment \cite{PhysRevLett.120.141101}, the common method to obtain constraint on individual coefficients is the method of the maximum reach analysis (MRA), in which one assumes that the scalar field in question only couples to one SM sector (one coupling parameter) and other coupling parameters are kept equal to zero, thus we can exclude one nonvanishing coupling coefficient at a time. More detailed information about this method can be found in Refs.\cite{PhysRevLett.117.241301,PhysRevLett.119.201102}.

\subsection{Constraints from $G$-measurement experiments }\label{sec33}

Considering $G$-measurement gravity experiments with torsion pendulum, the {scalar-field Yukawa} force leads to a variation in torque that is dependent on the {scalar field mass} $m_{\phi}$ and experimental design. Then, we analyzed the fluctuations of torque in the torsion pendulum experiments measuring the gravitational constant $G$. Huazhong University of Science and Technology (HUST) group has reported the $G$ measurements by two different methods (HUST-18), the time-of-swing (TOS) method and angular-acceleration-feedback (AAF) method \cite{li2018measurements}. The two methods obtained $G$ values of $6.674184(78)\times10^{-11}$ m$^{3}$kg$^{-1}$s$^{-2}$ and $6.674484(78)\times10^{-11}$ m$^{3}$kg$^{-1}$s$^{-2}$. In the absence of {a light scalar field}, the two measured values of gravitational constant $G$ should be same or their ratio should be equal to one. The insignificant difference between two $G$ values measured in one laboratory may be caused by the couplings of scalar field to the SM particles. By combining the $\mathcal{M}_{i}(m_{\phi},d_{i})$ in Eq.(\ref{dmg1}), we search {the possible scalar-field signal in these two experiments}.

In the TOS method, the experiment used pendulum to measure the changes in torsional oscillation frequency for two different source-mass configurations: near position and far position (the experimental schematic diagram can be found in Fig. 1 of Ref.\cite{li2018measurements}). The near configuration can speed up the oscillation and the far configuration can lead to a slower oscillation. The {Yukawa forces induced by the scalar fields} can produce a variation in the changes in torsional oscillation frequency for two configurations. Considering the interaction between the pendulum and source masses, the corresponding torques of the TOS experiment can be written as
\begin{equation}\label{dm8}
 \tau_{n/f}+ \tau_{\text{sf}n/f}=-k_{1n/f}\theta-k_{3n/f}\theta^3,
\end{equation}
where $k_{1n/f}=\partial^{2}V_{n/f}(\theta)/\partial^{2}\theta$, $k_{3n/f}=\partial^{4}V_{n/f}(\theta)/6\partial^{4}\theta$, $V_{n/f}(\theta,\alpha_{p},\alpha_{s})=-G\int[(1+\alpha_{p}\alpha_{s}e^{-m_{\phi}r})\rho_{p}\rho_{s}/r]dV_{p} dV_{s}$ is the gravitational potential energy between the pendulum and source masses that contains the {scalar-field} effects, $\rho_{p}$ is the mass distribution of pendulum, $\rho_{s}$ is the distribution of source masses, the subscripts $n$ and $f$ represent near and far source mass positions, respectively. Neglecting the nonlinear terms, the pendulum frequency squared for two configurations is given by $\omega^{2}_{n/f}=(k_{n/f}+k_{1n/f}(m_{\phi},\alpha_{p},\alpha_{s}))/I$ with $I$ the inertial moment of pendulum. As suggested in the TOS method \cite{heyl1930redetermination,PhysRevLett.102.240801,PhysRevD.82.022001,PhysRevLett.126.211101}, the term  $k_{1n/f}(m_{\phi},\alpha_{p},\alpha_{s}))$ can be treated as the effective gravitational torsion constant, which can be written as the form $G(C_{gn/f}+C_{\text{sf}n/f}(m_{\phi},\alpha_{p},\alpha_{s}))$, where $C_{gn/f}$ and $C_{\text{sf}n/f}(m_{\phi},\alpha_{p},\alpha_{s}))$ are the calculated parameter functions determined by the mass distributions of the pendulum and source masses in the apparatus. The TOS method gives
\begin{equation}\label{dm9}
  G(\Delta C_{g}+\Delta C_{\text{sf}}(m_{\phi},\alpha_{p},\alpha_{s}))=I\Delta \omega^{2}-\Delta k,
\end{equation}
where $\alpha_{p}$ is the {scalar-field} parameter of the pendulum, $\alpha_{s}$ is the {scalar-field} parameter of the source mass, $\Delta k$ represents the possible changes of spring constant for two configurations, the Newtonian gravity contribution is given by
\begin{equation}\label{nt1}
  \Delta C_{g}=C_{gn}-C_{gf}=-\frac{\partial^2}{\partial\theta^{2}}\int_{n} \frac{\rho_{p}\rho_{s}}{r}d^{3}r_{p}dr^{3}_{s} +\frac{\partial^2}{\partial\theta^{2}}\int_{f} \frac{\rho_{p}\rho_{s}}{r}d^{3}r_{p}dr^{3}_{s} ,
\end{equation}
the {scalar-field} contribution is
\begin{equation}\label{nt2}
  \Delta C_{\text{sf}}= C_{\text{sf}n}-C_{\text{sf}f}=-\frac{\partial^2}{\partial\theta^{2}}\int_{n} \frac{\alpha_{p}\alpha_{s}\rho_{p}\rho_{s}}{r}e^{-m_{\phi}r}d^{3}r_{p}dr^{3}_{s} +\frac{\partial^2}{\partial\theta^{2}}\int_{f} \frac{\alpha_{p}\alpha_{s}\rho_{p}\rho_{s}}{r}e^{-m_{\phi}r}d^{3}r_{p}dr^{3}_{s}.
\end{equation}
Based on the mass distributions of the pendulum and source masses, we can obtain the terms $\Delta C_{g}$ and $\Delta C_{\text{sf}}$ by the numerical integration in the laboratory coordinate system (see Appendix \ref{appe1}).

The schematic diagram and basic information of the TOS experiment can be found in Ref.\cite{li2018measurements}. The main parameters of the TOS experiment are listed in TABLE.\ref{tab:tos}.
The material of the pendulum is SiO$_{2}$. The material of source masses is SS316 stainless steel that is composed of a 62:18:14:3:2:1 ratio of Fe, Cr, Ni, Mo, Mn and Si numbers. The values for the dilaton charges in the TOS method are given in Table. \ref{tab:dm1}.
Using the Eq.(\ref{dm9}), the TOS experiment can be used to probe the possible {Yukawa effects mediated by scalar fields}. The gravitational constant $G$ can be eliminated by combining the measurement of AAF method.

\begin{table}[!t]
\caption{\label{tab:dm1} The dilaton charges for the materials in this work.}
\newcommand{\tabincell}[2]{\begin{tabular}{@{}#1@{}}#2\end{tabular}}
\begin{tabular}{lccccc}
\hline
\hline
\tabincell{l}
Experiment's \quad &$Q_{\hat{m}}$  \quad& $Q_{\delta m}$  \quad &$Q_{ m_e}$    \quad &$Q_{e}$\\
material    &$\times 10^{-2}$   &$\times 10^{-4}$    &$\times 10^{-4}$  &$\times 10^{-4}$   \\
\hline
  Wolfram               & 8.5   & 3.3   & 2.2    &42 \\
  Pendulum ($\text{SiO}_{2}$)  &8.0   & 0.02   & 2.8   & 15\\
  Source mass (SS316)   & 8.3    &1.2   &2.6     &26\\
  Pt/Rh (9:1)   & 8.5    &3.3   &2.2     &42\\
  Ti/Al/V (90:6:4)  & 8.3    &1.4   &2.5     &22\\
  Pt   & 8.5    &3.4   &2.2     &43\\
  Cu   & 8.3    &1.3   &2.5     &27\\
  Al   & 8.1    &0.6   &2.6     &17\\
\hline
\hline
\end{tabular}
\end{table}

In the AAF method, the experiment used two turntables to rotate the torsion pendulum coaxially and source masses individually (the experimental schematic diagram can be found in Fig. 1 of Ref.\cite{li2018measurements} and the main parameters of the AFF experiment are listed in TABLE.\ref{tab:tos}). The pendulum experienced a sinusoidal torque caused by the interactions of the source masses. A high-gain feedback control system was used to reduce the twist angle of the fiber to about zero. The resulting angular acceleration of the pendulum equals the angular acceleration of gravity and {scalar-field} force generated by the source masses,
\begin{equation}\label{dm10}
I\iota(t)=(\tau+\tau_{\text{sf}})\sin(\omega_{s}t),
\end{equation}
where $\iota(t)$ is the angular acceleration of the inner turntable and $\omega_{s}$ is the signal frequency. Through Fourier transform and considering the inertia torque in the signal frequency $\omega_{s}$, the corresponding torque is given by $\tau+\tau_{\text{sf}}=I\iota(\omega_{s})$.
As suggested in the AAF method \cite{PhysRevLett.85.2869,luo2013thermal,xue2014preliminary}, the torque terms $\tau$ and $\tau_{\text{sf}}$ can be treated as the gravitational coupling constants $GD_{g}$ and $GD_{\text{sf}}(m_{\phi},\alpha_{p},\alpha_{s})$, which are determined by the mass distributions of the pendulum and source masses in the AAF apparatus. The AAF method gives
\begin{equation}\label{dmaas}
  G(D_{g}+D_{\text{sf}}(m_{\phi},\alpha_{p},\alpha_{s}))=\iota(\omega_{s}).
\end{equation}
In this equation, the Newtonian term is determined by
\begin{equation}\label{dmn1}
  D_{g}=\frac{-8\pi}{I} \sum^{\infty}_{l=2}\frac{1}{2l+1}\sum^{l}_{m=0}m q_{lm} Q_{lm},
\end{equation}
where $q_{lm}=\int\rho_{p}(r_{p})Y^{m*}_{l}(\theta_{p},\varphi_{p})r^l_{p}d^3 r_{p}$ is the Newtonian multipole moments of pendulum,
and $Q_{lm}=\int\rho_{s}(r_{s})\times$ $Y^{m}_{l}(\theta_{s},\varphi_{s})r^{-(l+1)}_{s} d^3 r_{p}$ is the Newtonian multipole moments of source mass.
The {scalar-field contribution} is given by using a similar calculation
\begin{equation}\label{dm11}
  D_{\text{sf}}(m_{\phi},\alpha_{p},\alpha_{s})=\frac{-8\pi}{I} \sum_{l=2}^{\infty}\frac{1}{2l+1}\sum^{l}_{m=0}mq^{\text{sf}}_{lm}Q^{\text{sf}}_{lm},
\end{equation}
where
\begin{equation}\label{dm12}
  q^{\text{sf}}_{lm}=\alpha_{p}\int\rho_{p}(r_{p})\frac{(2l+1)!!}{(m_{\phi})^{l}}i_{l}(m_{\phi}r_{p})Y^{m*}_{l}(\theta_{p},\varphi_{p})d^3 r_{p}
\end{equation}
is the mass multipole moments of the pendulum induced by the {scalar fields}, and
\begin{equation}\label{dm13}
  Q^{\text{sf}}_{lm}=\alpha_{s}\int\rho_{s}(r_{s})\frac{(m_{\phi})^{l+l}}{(2l-1)!!}k_{l}(m_{\phi}r_{s})Y^{m}_{l}(\theta_{s},\varphi_{s})d^3 r_{s}
\end{equation}
is the mass multipole moments of source mass induced by the {scalar fields}. $i_{l}(m_{\phi}r_{p})$ and $k_{l}(m_{\phi}r_{s})$ are the spherical modified Bessel functions. The terms $D_{g}$ and $D_{\text{sf}}(m_{\phi},\alpha_{p},\alpha_{s})$ were calculated by the numerical integration in the laboratory coordinate system (see Appendix \ref{appe1}).
The schematic diagram and main experimental information of the AAF method can be found in Ref.\cite{li2018measurements}. The AAF experiment used the pendulums and source masses of the same materials as the TOS experiment. The values of dilaton charges are the same as those of the TOS method.

The separations between the pendulum and source masses are different for AAF and TOS methods so they are sensitive to different field masses. From the basic parameters of the experiments in TABLE.\ref{tab:tos}, we can estimate the effective separations for the two methods. The effective separation can be estimated from the experimental parameters, which is slightly less than the separation between geometric centers (GCs) of source mass and pendulum or half the separation between GCs of two source masses, and slightly greater than the separation between GC of source mass and boundary of pendulum. Therefore, we can obtain the effective separations of the TOS and AFF experiments as about $5-8$ cm and $10-20$ cm, respectively. It demonstrates that the HUST G-measurement experiments are most sensitive to the separation ranges of centimeters and decimeters. The ISL experiments are sensitive to the millimeter and submillimeter ranges. The gravitational-wave interferometers are sensitive to ranges greater than $10^{2}$ m. Therefore, for the parameter constraints, the results of $G$-measurement experiments can complement that of ISL experiments and gravitational-wave interferometers.
Comparing these two measurements, one can obtain a $G$value-independent quantity
\begin{equation}\label{dmc11}
  \frac{1+D_{\text{sf}}/D_{g}}{1+\Delta C_{\text{sf}}/\Delta C_{g}}
  =\frac{I_{\text{A}}\iota(\omega_{s})/D_{g}}{(I_{\text{T}}\Delta \omega^{2}-\Delta k)/\Delta C_{g}}=\frac{G_{\text{AAF}}}{G_{\text{TOS}}},
\end{equation}
where $I_{A}$ and $I_{T}$ are the inertial moments of the AAF and TOS pendulums, respectively. In the absence of {the scalar field}, this equation recovers
the results of experimental $G$ measurements. The terms $I_{\text{A}}\iota(\omega_{s})/D_{g}$ and $(I_{\text{T}}\Delta \omega^{2}-\Delta k)/\Delta C_{g}$ are equal to the measurements of $G_{\text{AAF}}$ and $G_{\text{TOS}}$, respectively. We consider the experiments of the AAF-II and TOS-I-Fibre 3 \cite{li2018measurements}, where the $G$ values are $G_{\text{AAF}}=6.674375(164)\times10^{-11}$  m$^{3}$kg$^{-1}$s$^{-2}$ and $G_{\text{TOS}}=6.674269(183)\times10^{-11}$ m$^{3}$kg$^{-1}$s$^{-2}$ with two standard deviations. Using the calculated parameters $D_{\text{sf}}(m_{\phi},d_{i})$ and $\Delta C_{\text{sf}}(m_{\phi},d_{i})$, the coupling coefficients can be limited from the Eq.(\ref{dmc11}).

\begin{figure}
\includegraphics[width=0.5\textwidth]{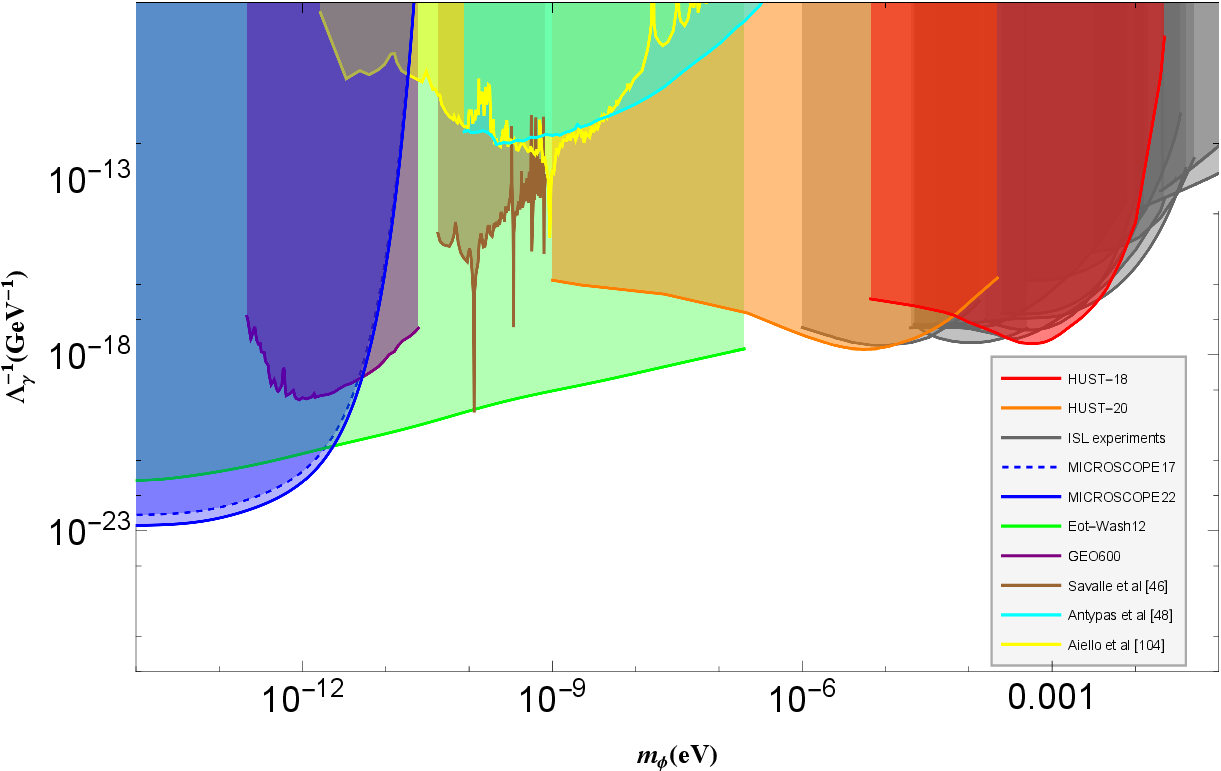}
\includegraphics[width=0.5\textwidth]{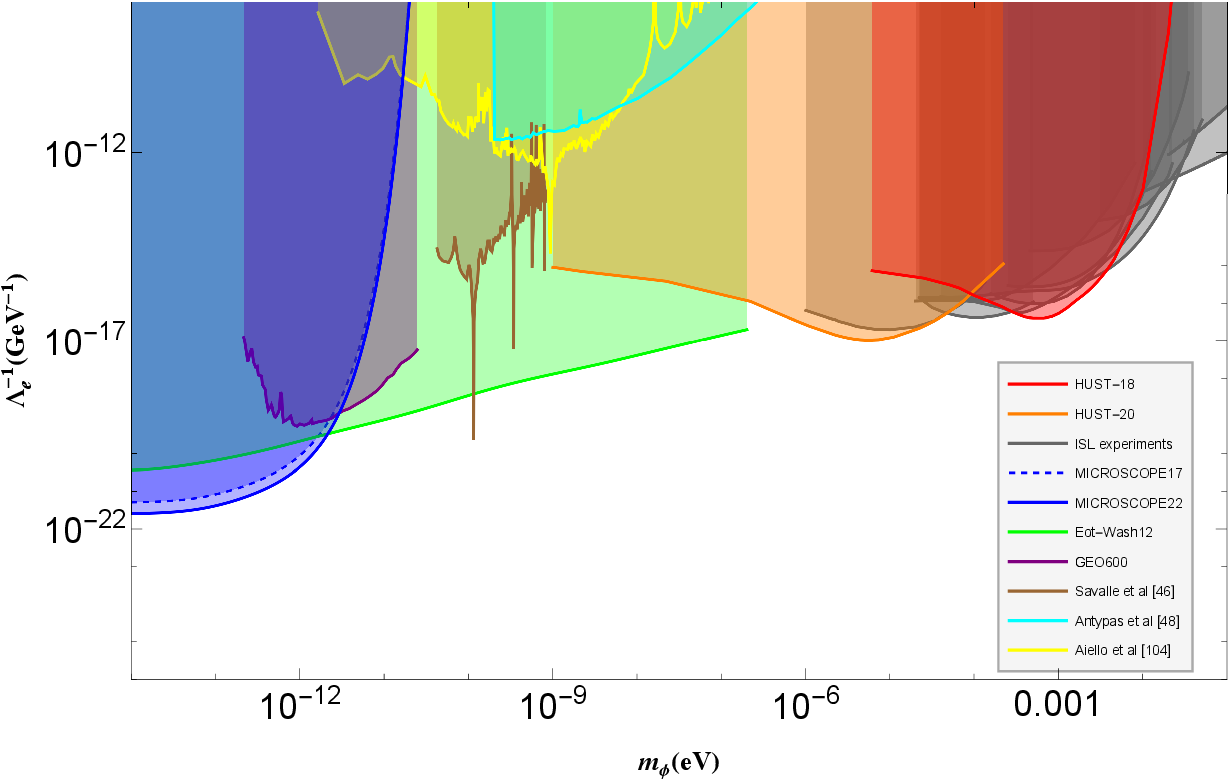}
\caption{\label{fig:a2} Constraints on the coupling parameters $\Lambda_{\gamma}$ (upper) and $\Lambda_{e}$ (lower) as a function of the mass of the scalar field $m_{\phi}$. The colored regions represent the constraints on photon coupling $\Lambda_{\gamma}$ and electron coupling $\Lambda_{e}$ at the $95\%$ confidence level. The red region indicates the limited parameter spaces for the coupling parameters by our result in the HUST torsion pendulum experiments. The blue excluded regions mark the parameter spaces from the first result \cite{PhysRevLett.120.141101} ($MICROSCOPE17$) and final result ($MICROSCOPE22$) \cite{PhysRevLett.129.121102} of $MICROSCOPE$ experiment. The green excluded regions are from the E$\ddot{\text{o}}$t-Wash WEP test \cite{PhysRevLett.100.041101,wagner2012torsion}. The gray excluded regions represent the short-range ISL experiments \cite{PhysRevLett.86.1418,PhysRevLett.124.101101,long1999experimental,long2003upper,PhysRevLett.98.021101,PhysRevLett.107.171101,PhysRevLett.116.221102,PhysRevLett.90.151101,PhysRevLett.98.201101,
PhysRevLett.108.081101,PhysRevLett.116.131101,PhysRevD.32.3084,PhysRevD.70.042004,PhysRevD.78.022002}.
Other colored regions denote parameter spaces excluded by the previous direct searches in experiments, including GEO600 interferometer (purple) \cite{vermeulen2021direct}, Savalle et al. (brown) \cite{PhysRevLett.126.051301}, Antypas et al. (cyan) \cite{PhysRevLett.123.141102}, and Aiello et al. (yellow) \cite{PhysRevLett.128.121101}.}
\end{figure}

Using MRA method, we use the HUST-18 experiment and Eq.(\ref{dmc11}) to set the constraints on the {scalar-field coupling}  parameters at a 95$\%$ confidence level (We focus on the constraint on the coefficients $\Lambda_{\gamma}$ and $\Lambda_{e}$, and other coefficients can be limited by the same way.).
The electron coupling $\Lambda_{e}$ and photon coupling $\Lambda_{\gamma}$ parameters are constrained as a function of the scalar field's mass $m_{\phi}$ (For a comparison with the previous results, the relation between the two sets of coefficients is $\Lambda_{\gamma}={M_{\text{Pl}}}/({\sqrt{4\pi}d_{e}})$ and $\Lambda_{e}={M_{\text{Pl}}}/({\sqrt{4\pi}d_{m_{e}}}$)).
The constraints in the mass range $10^{-9}-10^{-4}$ eV from our analysis are plotted with orange excluded regions in Fig. \ref{fig:a2}, together with previously published upper limits.
Up to now, the most stringent constraints on the scalar-field coupling parameters with low masses ranges are given by the $MICROSCOPE$ experiment \cite{PhysRevLett.120.141101,PhysRevLett.129.121102} and Be/Ti torsion pendulum experiment of E$\ddot{\text{o}}$t-Wash group \cite{PhysRevLett.100.041101,wagner2012torsion}.
The gravitational-wave detectors (or interferometers) have recently seen a strong surge thanks to their excellent sensitivity at or beyond quantum limits. The gravitational-wave detectors GEO600, LIGO and Virgo set a strong limit on the coupling coefficients $\Lambda_{\text{e}}$ and $\Lambda_{\gamma}$ for the masses range of $10^{-13}-10^{-11}$ eV \cite{vermeulen2021direct,PhysRevD.105.063030,guo2019searching}. Larger masses of $10^{-12}-10^{-7}$ eV were constrained by the interferometer experiments \cite{PhysRevLett.126.051301,PhysRevLett.128.121101}.
HUST-18 experiments set new limits on {coupling coefficients} at up to $\Lambda_{\text{e}}=1\times10^{17}$ GeV and $\Lambda_{\gamma}=7\times10^{17}$ GeV for mass between $10^{-9}$ and $10^{-4}$ eV.
Compared with the current limits in the mass range $10^{-9}-10^{-7}$ eV from the interferometers, HUST-18 constraints improve by more four orders of the magnitude. Different from E$\ddot{\text{o}}$t-Wash EP experiments and $MICROSCOPE$ experiment, our results are more competitive for the scalar field masses larger than a few $10^{-6}\text{eV}$.

\subsection{Constraints from Inverse-Square Law test experiments }\label{sec31}
Here we consider the constraints on {the scalar field} parameter space from the short-range ISL experiments in laboratory. HUST-20 ISL test experiment was the torsion balance experiments that tested Newton ISL in the ranges of submillimeter. In this ISL test experiment, we analyzed fluctuations of the torque at a modulating signal frequency of HUST-20 torsion pendulum experiment, and the schematic drawing of the experimental setup is shown in Fig.1 of Ref.\cite{PhysRevLett.124.051301}. The experiment adopted the dual-modulation method and dual-compensation technology. The dual-modulation method was used to separate the signal and disturbance frequencies, and to reduce the torque noise. The attractor was eightfold azimuth symmetrically distributed and rotated around a horizontal axis with the driving frequency $\omega_{d}=1.634$ mrad/s. The signal frequency was set at $8 \omega_{d}$, which effectively separates the disturbances of the fundamental frequency. By adding corresponding compensation masses on both the pendulum and attractor, the dual-compensation design can realize a null experiment for the Newtonian torque at $8 \omega_{d}$. Considering the exponential force, this design does not suppress the changes in the Yukawa-like torque. Thus, the torque induced by scalar field can be detected with the signal frequency $8 \omega_{d}$ in the most sensitive direction of the torsion balance. More details about the experimental setup can be in the Refs.\cite{PhysRevLett.124.051301,PhysRevLett.116.131101}.
From the interaction of Eq.(\ref{dm3}), the equation of motion for the closed-loop torsion pendulum is given by
\begin{equation}\label{dm6}
  I\ddot{\theta}+k(1+i/Q)\theta-k_{e}\theta=\tau+\tau_{\text{sf}}-\beta U,
\end{equation}
where $\tau_{\text{sf}}(m_{\phi},d_{a})$ represents the torque signal of {scalar fields} that is given by
\begin{equation}\label{dm6dm1}
  \tau_{\text{sf}}(m_{\phi},d_{a})=\frac{\partial}{\partial\theta}\int
  \frac{G\rho_{i}\rho_{j}}{r}\alpha_{A}\alpha_{B}e^{-m_{\phi}r}d^{3}r_{i}d^{3}r_{j},
\end{equation}
$I$ is the inertial moment of the pendulum, $k$ is the spring constant, $Q$ is the quality factor, $k_{e}$ is the negative spring constant induced by the electrostatic interaction, $\tau$ represents torque in the Newtonian gravity frame, $\beta$ is the ratio of the control torque to feedback voltage $U$. The systematic uncertainty and errors can be found in the Ref.\cite{PhysRevLett.124.051301}.

The experiments were performed at four separations, including 210, 230, 295, and 1095 $\mu$m. The residual Newtonian torques are not more than $1.5\times10^{-17}$ Nm for these four measurements, which are considered in all experiments. The measured data can be split into four data subsets.
For constraining the coupling parameters of the scalar fields, the maximum likelihood estimate method was used. The likelihood function is
\begin{equation}\label{dm7}
  P(\tau_{\text{sf}},\tau_{m},d_{a},m_{\phi})=\prod_{i}\frac{1}{\sqrt{2\pi} \sigma_{i}}e^{-[(\tau_{mi}-\tau_{\text{sf}i})^2/2\sigma_{i}^2]},
\end{equation}
where $i$ denotes the different separation experiments, $\tau_{m}$ is the in-phase component of the measured torques, $\sigma_{i}$ is the total error of $\tau_{m}$ and $\tau_{\text{sf}}$, and $\tau_{\text{sf}}(m_{\phi},d_{a})$ is the possible {scalar-field} torque at $8 \omega_{d}$, which is calculated by numerical integration from the geometric parameters (see Appendix \ref{appe2}). The experiments were performed with the tungsten test mass and attractor. We have the parameter $\alpha_{W}$ for the {scalar-field} torque (as shown in Table \ref{tab:dm1}). Combining four data subsets, we use the likelihood function (\ref{dm7}) to set the constraints on the coupling parameters as a function of the mass of the field $m_{\phi}$. For the limits on the coupling parameters with 95$\%$ confidence level, the likelihood function is characterized by $(1/C)\int P(\tau_{\text{sf}},\tau_{m},d_{a},m_{\phi}) d d_{a}=95\%$, where $C$ is the normalization coefficient.

Using the MRA method, we use the HUST-20 ISL test experiment to set the constraints on the electron coupling $\Lambda_{e}$ and photon coupling $\Lambda_{\gamma}$ parameters as a function of the scalar field's mass $m_{\phi}$.
Our limits from the HUST-20 ISL test experiment are plotted with red lines in Fig.\ref{fig:a2}.
HUST-18 experiments set the limits on {coupling parameters} at up to $\Lambda_{\text{e}}=3\times10^{16}$ GeV and $\Lambda_{\gamma}=5\times10^{17}$ GeV for {mass ranges} between $10^{-5}$ and $10^{-1}$ eV.
These results represent an important contribution to a largely unexplored region of the scalar field parameter space.
The short-range ISL experiments in laboratories are sensitive to the mass range parameter spaces. The tests of gravitational ISL have been performed by University of Washington, Stanford, IUPUI, HUST, Colorado, Irvine, Yale among others, and many remarkable results have been reported these groups \cite{PhysRevLett.86.1418,PhysRevLett.124.101101,long1999experimental,long2003upper,PhysRevLett.98.021101,PhysRevLett.107.171101,PhysRevLett.116.221102,
PhysRevLett.90.151101,PhysRevLett.98.201101,PhysRevLett.108.081101,PhysRevLett.116.131101,PhysRevD.32.3084,PhysRevD.70.042004,PhysRevD.78.022002}, which can translate into the constraints on the {scalar field} parameter spaces. In fig.\ref{fig:a2}, the gray excluded regions present the constrains from these short-range ISL experiments. For example, Hoyle \emph{et al}. high-precision tested the gravitational ISL at submillimeter ranges\cite{PhysRevLett.86.1418}. A rough analysis demonstrates that this ISL experiment can set strong limits on the {scalar field} parameter spaces and extend the limits slightly at the lighter mass region. Lee \emph{et al}. reported a highly accurate test of the gravitational ISL at separations down to 52 $\mu$m \cite{PhysRevLett.124.101101}. A rough analysis demonstrates that this experiment can set stringent constraints on the {scalar field} parameter spaces and extend the limits slightly at the higher mass region. Other short-range ISL experiments also can set strong excluded regions on the corresponding mass ranges. All these experimental bounds complement each other and represent an important torsion-balance experiment contribution to the possible searches of {the long-range Yulawa forces  mediated by scalar fields.}

\begin{figure}
\includegraphics[width=0.5\textwidth]{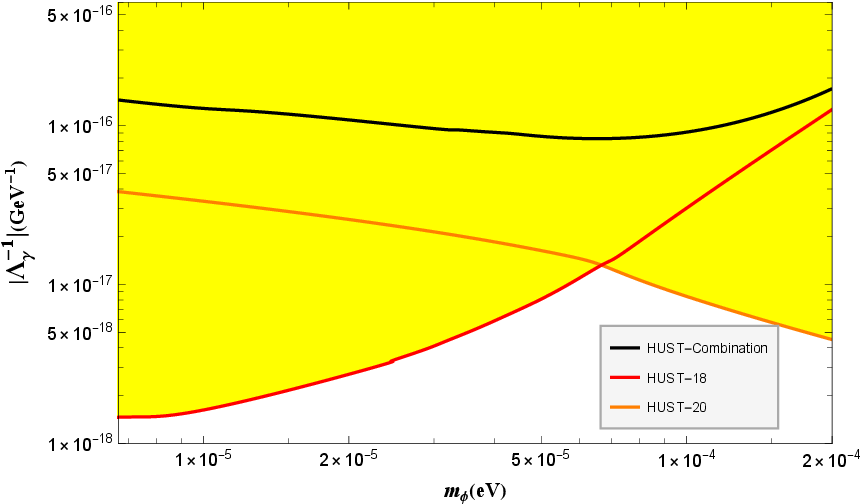}
\caption{\label{fig:3} The nonMRA constraints on the photon coupling parameters $|\Lambda_{\gamma}|$ as a function of the mass of the scalar field $m_{\phi}$. The black line represents the limits from the combination of HUST-18 and HUST-20 experiments. }
\end{figure}

\begin{figure}
\includegraphics[width=0.5\textwidth]{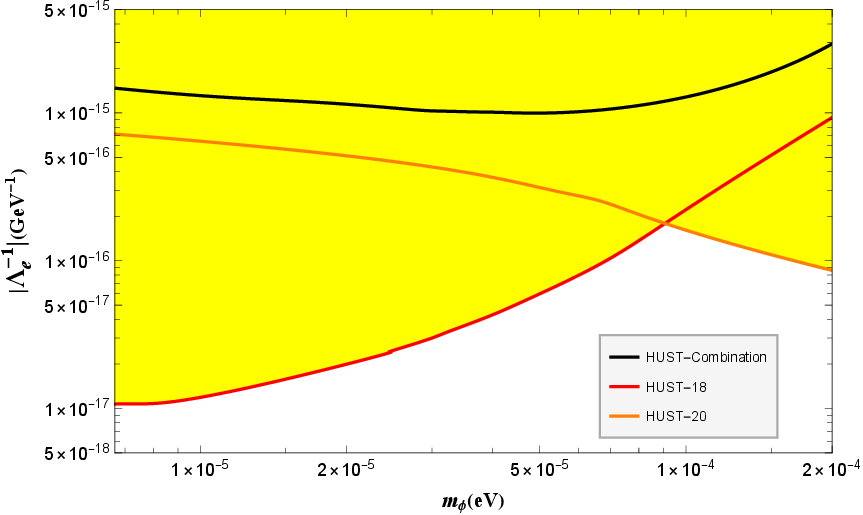}
\caption{\label{fig:4} The nonMRA constraints on the electron coupling parameters $|\Lambda_{e}|$ as a function of the mass of the scalar field $m_{\phi}$. }
\end{figure}

The MRA method obtains an idealistic estimate since it assumes that the experiment is sensitive to only one coupling parameter. By using multiple sets of data with different combinations of the coupling coefficients, we can set the constraints on the individual coefficients without the assumption that the values of other parameters are zero. In the mass ranges $5\times10^{-6}-2\times10^{-4}$ eV, HUST-18 and HUST-20 experiments are sensitive to the different combinations of the coupling coefficients,
\begin{eqnarray}\label{nmra}
% \nonumber % Remove numbering (before each equation)
  \Delta G_{18} &=&\alpha_{s}\alpha_{p} \Delta G_{\text{sf}}(m_{\phi}) =\frac{M^{2}_{\text{Pl}}}{4\pi }\left( \frac{1.5\times10^{-3}}{\Lambda_{\lambda}}+\frac{2.8\times10^{-3}}{\Lambda_{e}}\right)\left(\frac{2.6\times10^{-3}}{\Lambda_{\lambda}}+\frac{2.6\times10^{-3}}{\Lambda_{e}}\right)\Delta G_{\text{sf}}(m_{\phi}), \\
  \tau_{20}&=& \alpha_{W}\alpha_{W} \tau_{\text{sf}}(m_{\phi}) =\frac{M^{2}_{\text{Pl}}}{4\pi }\left( \frac{4.2\times10^{-3}}{\Lambda_{\lambda}}+\frac{2.2\times10^{-3}}{\Lambda_{e}}\right)\left(\frac{4.2\times10^{-3}}{\Lambda_{\lambda}}+\frac{2.2\times10^{-3}}{\Lambda_{e}}\right) \tau_{\text{sf}}(m_{\phi}),
\end{eqnarray}
where the $\Delta G_{\text{sf}}(m_{\phi})$ and $\tau_{\text{sf}}(m_{\phi})$ are the numerical results of {scalar-field} effects in the HUST-18 and HUST-20 experiments, respectively. By using two sets of HUST-18 and HUST-20 data, one can obtain the solutions ($\Lambda_{\gamma}, \Lambda_{e}$) for each value of $m_{\phi}$. Considering the max one of absolute values ($|\Lambda_{\gamma}|, |\Lambda_{e}$|), we can set final exclusion limits on the $|\Lambda_{\gamma}|$ and $|\Lambda_{e}|$ at 95$\%$ confidence level, which does not rely on MRA method. Figures \ref{fig:3} and \ref{fig:4} present the excluded regions for the photon coupling parameter $\Lambda_{\gamma}$ and electron coupling parameter $\Lambda_{e}$, respectively.
Although these limits have less sensitivity than those gained by the MRA method, the results do not depend on the MRA assumption.
Better limits can be achieved by combining more $G$-measurement or ISL test experiments with different materials.

{It is interesting to search in a mass range with multiple gravity experiments.} The combination of multiple sets of data can be used to constrain the {coupling parameters} without the MRA assumption. By reducing electrostatic force, further improvements could be made in the HUST gravitational ISL experiment. The ISL test experiments with different materials also {can be performed to provide a new data set for constraining coupling parameters.} And more $G$-measurement experiments will be considered using Eq.(\ref{dmg1}) in future work. Combining multiple short-range gravity experiments, we can obtain more stringent constraints on the coupling parameters in a larger mass range.

\subsection{Constraints from equivalence principle experiments}\label{sec32}
Considering the measurements of the weak equivalence principle, the constraints on the E\"{o}tv\"{o}s parameter can be directly translated into the bounds on {the coupling parameter spaces} $(D_{\hat{m}},D_{e},m_{\phi})$ or $(d_{a},m_{\phi})$. Currently, the most accurate measurement is given by the final result of the $MICROSCOPE$ mission \cite{PhysRevLett.129.121102}  $\eta=(-1.5\pm5.5)\times10^{-15}$ at a 2$\sigma$ confidence level, which can provide stronger constraints than previous results on the {the coupling parameter spaces}. From Eq.(\ref{epdk6}), the final result of the $MICROSCOPE$ mission can be expressed as
\begin{equation}\label{ceq1}
  \eta({\text{Pt,Ti}})=\left[D_{\hat{m}}\left( Q^{*}_{\hat{m}}({\text{Pt}})- Q^{*}_{\hat{m}}({\text{Ti}})\right) +D_{e} \left( Q^{*}_{e}({\text{Pt}})- Q^{*}_{e}({\text{Ti}}) \right) \right]I^{(n)}_{\text{eff}}(m_{\phi},R_{E})
\end{equation}
where the dilaton charges are given by Table.{\ref{tab:dm1}}. {After giving the mass of a scalar field}, it is convenient to obtain the parameter spaces $(D_{\hat{m}},D_{e},m_{\phi})$. Figure. \ref{fig:cp1} show the allowed $(D_{\hat{m}},D_{e},m_{\phi})$ parameter space of the $MICROSCOPE$ final result. $MICROSCOPE$ is sensitive to masses in the ranges $m_{\phi}\lesssim 1\times10^{-13}$ eV, and it loses some sensitivity for larger masses ranges.

In addition, by using the MRA method, one can set the constraints on the individual coupling coefficients. Focusing on the photon and electron couplings, one translates the constraints on the parameter $\eta$ into bounds on the coupling parameters $\Lambda_{\gamma}$ and $\Lambda_{e}$.
From Eq.(\ref{epdk8}), the final measurement of $MICROSCOPE$ leads to the constraint on the parameter $\Lambda_{a}$ as
\begin{equation}\label{ceq2}
  \Lambda_{a}=M_{\text{Pl}}\sqrt {\frac{\Delta Q_{a} Q_{a,E}I^{(n)}_{\text{eff}}({m_{\phi}}R_{E})}{4\pi\eta({\text{Pt,Ti}})}}.
\end{equation}
In  Fig.\ref{fig:a2}, the blue regions represent the excluded parameter spaces on the coupling parameters $\Lambda_{\gamma}$ and $\Lambda_{e}$ from the $MICROSCOPE$ results.
The blue dashed lines and solid lines are the parameter constraints through the first result ($MICROSCOPE17$) and final result ($MICROSCOPE22$) of the $MICROSCOPE$ mission, respectively. The $MICROSCOPE22$ results set the constraints on {coupling parameters} at up to $\Lambda_{\gamma}=7\times10^{22}$ GeV and $\Lambda_{e}=4\times10^{21}$ GeV for the mass ranges $m_{\phi}\lesssim 10^{-13}$ eV. It allows to exclude corresponding regions above $|d_{e}|=4\times10^{-5}$ and $|d_{m_{e}}|=8\times10^{-4}$ for the mass ranges $m_{\phi}\lesssim 10^{-13}$ eV. Considering WEP test instrument 1 m above the ground and $n$-layers Earth model, the E$\ddot{\text{o}}$t-Wash WEP test can be used to set stringent bounds on parameter spaces below the mass ranges $10^{-7}$ eV  \cite{PhysRevLett.100.041101,wagner2012torsion}.

\begin{figure}
\includegraphics[width=0.5\textwidth]{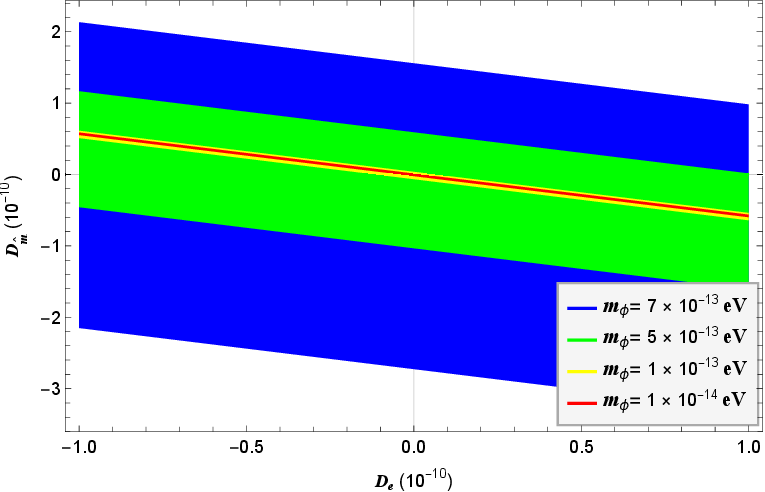}
\caption{\label{fig:cp1} The constraints on {the coupling parameter spaces} $(D_{\hat{m}},D_{e},m_{\phi})$ of the $MICROSCOPE$ final result. The Blue, green, yellow and red colors represent the allowed regions for the {scalar field} masses of $7\times10^{-13}$ eV, $5\times10^{-13}$ eV, $1\times10^{-13}$ eV and $1\times10^{-14}$ eV, respectively.  }
\end{figure}

\section{Conclusion}\label{sec4}

In conclusion, we have presented {the searches for long-range Yukawa forces mediated by light scalar fields} with three types of gravity experiments: the big-$G$ measurement experiments, Newton ISL test experiments, and equivalence principle test experiments. By analyzing several existing gravity experiments using the MRA method, we have set the experimental excluded regions for the {scalar field coupling to the standard model particles} within large mass ranges.
In the $G$-measurement experiments, we proposed a method to constrain the {scalar field} by utilizing two or multiple $G$-measurement experiments.
With the two $G$-value measurements of HUST-18 $G$-measurement torsion-balance experiments, we have placed the bounds on the {coupling parameters} at up to $\Lambda_{\gamma}=7\times10^{17}$ GeV and $\Lambda_{e}=1\times10^{17}$ GeV in the mass range $10^{-9}-10^{-4}$ eV. This result demonstrates a successful application of $G$-measurement gravity experiments to bound the {coupling parameters of the scalar fields}.
In the ISL experiments, we studied the {scalar-field} influences in the orbital motions of planets or satellites, and the short-range ISL test experiments. From the HUST-20 ISL test torsion-balance experiment, we have obtained limits on the photon coupling and electron coupling at up to $\Lambda_{\gamma}=5\times10^{17}$ GeV and $\Lambda_{e}=3\times10^{16}$ GeV in the mass ranges $10^{-5}-10^{-1}$ eV. These results represent an experimental contribution to an unexplored mass region of {the parameter space of light scalar fields}.
In the equivalence principle experiments, a $n$-layers Earth model is developed to be used to set limits on the {parameter space} from the measurements of E\"{o}tv\"{o}s parameter.
By analyzing the final result of $MICROSCOPE$ mission, we have updated the constraints on the {coupling parameters} at up to $\Lambda_{\gamma}=7\times10^{22}$ GeV and $\Lambda_{e}=4\times10^{21}$ GeV for mass ranges $m_{\phi}\lesssim 10^{-13}$ eV, which updates the previous result of $MICROSCOPE$ mission. Furthermore, combining the HUST-18 and HUST-20 results, we also presented the non-MRA constraints on the photon coupling and electron coupling parameters, which are not based on maximum reach analysis. At present, the $G$-measurement experiments and ISL experiments are still ongoing in many groups, which may provide multiple sets of data. This may not only improve the non-MRA constraints on $\Lambda_{\gamma}$ and $\Lambda_{e}$, but also constrain multiple coefficients simultaneously without the assumption other coefficients are suppressed.

{The existence of a light scalar field provides a possible source for the violation of the weak equivalence principle or dark matter. The long-range interactions medicated by the light scalar fields have been constrained by various experiments and it is still important to diversify experimental efforts. The addition of $G$-measurement gravity experiments to this field is valuable. The combined analysis of gravity experiments, such as $G$-measurement experiments, ISL test experiments, or equivalence principle test experiments, may provide indispensable complementary information for the scalar long-range forces.
In the future, the improved limits on the scalar-field parameter space may be achieved by improving the precision of the torsion pendulum experiments through reducing electrostatic force or using new processing techniques, or by conducting the experiments with different materials of source masses since it can provide different sensitivities to the coupling parameters.}

\section{Acknowledgment}
The authors thank the anonymous referees for the helpful comments on this paper.
This work is supported by the National Natural Science Foundation of China (Grants No.12247150, No.12175076, No.11925503 and No.12150012), and the Post-doctoral Science Foundation of China (Grant No.2022M721257).

\begin{appendices}
\appendix
\section{The effect of the {scalar-field} force on the orbit motion}\label{appe3}

We consider the {Yukawa force induced by the light scalar fields} as a perturbation to the orbit motion. The unperturbed Kepler orbit is given by $r=a(1-e^{2})/(1+e\cos{f})$, where $a$ is the semimajor axis, $e$ is the eccentricity, $f$ is the true anomaly. To calculate the change of Kepler orbit, the disturbing acceleration is decomposed by the projections of radial $\mathcal{A}$, along-track $\mathcal{B}$ and cross-track directions $\mathcal{C}$.
By using the standard method of the Gauss perturbation equations, the changes of Kepler parameters can be expressed in terms of the projections $\mathcal{A}$, $\mathcal{B}$ and $\mathcal{C}$.
The Gauss equations are expressed as \cite{haranas2011yukawa}
\begin{equation}\label{fif2}
  \frac{da}{dt}=\frac{2}{n\sqrt{1-e^{2}}}\left[ e\cos{f}\mathcal{A}+\left(\frac{p}{r}\right)\mathcal{B}\right]
\end{equation}

\begin{equation}\label{fif3}
  \frac{d e}{d t}=\frac{\sqrt{1-e^{2}}}{na}\left[ \sin{f}\mathcal{A}+\left[ \cos{f}+\frac{1}{e}\left(\frac{a-r}{a}\right)\right]\mathcal{B}  \right]
\end{equation}

\begin{equation}\label{fif4}
  \frac{d i}{d t}=\frac{1}{na\sqrt{1-e^{2}}}\frac{r}{a}\cos{(\omega +f)}\mathcal{C},
\end{equation}

\begin{equation}\label{fif5}
  \frac{d \Omega}{dt}=\frac{1}{na\sin{i}\sqrt{1-e^2}}\frac{r}{a}\sin(\omega +f)\mathcal{C}
\end{equation}

\begin{equation}\label{fif6}
  \frac{d \omega}{dt}=\frac{\sqrt{1-e^{2}}}{nae}\left[-\cos{f}\mathcal{A}+\left(1+\frac{r}{p} \right)\sin{f}\mathcal{B}  \right]-\cos{i}\frac{d \Omega}{dt}
\end{equation}

\begin{equation}\label{fif7}
  \frac{d M}{dt}=n-\frac{2r}{na^{2}}-\sqrt{1-e^{2}}\left(\frac{d \omega}{dt}+\cos{i}   \frac{d \Omega}{dt}\right)
\end{equation}
where other Kepler parameters are given by the orbital inclination $i$, longitude of the ascending node $\Omega$, argument of perigee $\omega$, mean anomaly $M$, orbit semi-latus rectum $p=a(1-e^2)$ and Keplerian mean motion $n=\sqrt{GM_{S}/a^3}$. The orbital evolutions can be derived from the above Gauss equations. From the disturbing acceleration (\ref{fif1}), {the disturbing force} has only the radial component $\mathcal{A}$ affecting in-plane orbital motions. Thus, the orbital inclination $i$ and the longitude of the ascending node $\Omega$ are not affected by {the disturbing force}. Generally, the actual observables and measurements are from the long-term variations of the Kepler orbit parameter. By averaging the above Gauss equations over an orbital evolution, the secular variations of Keplerian parameters can be obtained.

\section{ The calculation on {scalar-field} parameters for the HUST-18 torsion-pendulum experiment}\label{appe1}
In the TOS and AAF experiment, the source masses are the spherical stainless steel bodies, which can be treated as uniform-density spherically symmetric bodies. For a uniform-density spherically symmetric body $A$ with the mass $M_{A}$ and radius $R_{A}$, the integration from {the Yukawa potential mediated by the scalar fields} leads to \cite{adelberger2003tests}
\begin{equation}\label{mp1}
  U_{\text{sf},A}(\textbf{\emph{x}})=\alpha_{A}I\left(m_{\phi}R_{A}\right)\frac{GM_{A}}{r}e^{-m_{\phi}r}
\end{equation}
with
\begin{equation}\label{mp2}
  I(x)=3\frac{x\cosh{x}-\sinh{x}}{x^{3}}.
\end{equation}

\begin{table}[!t]
\caption{\label{tab:tos} The main experimental parameters of the TOS experiment and AAF experiment \cite{li2018measurements}.}
\newcommand{\tabincell}[2]{\begin{tabular}{@{}#1@{}}#2\end{tabular}}
\begin{tabular}{lcccc}
\hline
\hline
\tabincell{l}
Experimental parameters   \qquad\qquad\qquad\qquad\qquad \,\,\,\,\,\,\,\, \,\,\,\,\,\,\,\,&TOS values    \qquad\qquad\,\,\,\,\,\,\,\, \,\,\,\,\,\,\,\, &AAF values   \\
\hline
  Pendulum:              &   &\\
  \quad Length                 & 91.0 mm   & 91.1 mm   \\
  \quad Width                  & 11.1 mm    & 4.0 mm \\
  \quad Height                 & 30.7 mm   & 49.9 mm \\
  \quad Mass                   & 68.1 g   & 40.0 g   \\
  Source Masses:         &   &\\
  \quad Mass of sphere no.2             & 778.2 g & \\
  \quad Mass of sphere no.4             & 778.0 g &\\
  \quad Diameter of sphere no.2         & 57.2 mm &\\
  \quad Diameter of sphere no.4         & 57.2 mm & \\
  \quad Horizontal distance of GCs 2-4       &157.2 mm &\\
  \quad Mass of sphere no.7             & & 8543.6 g\\
  \quad Mass of sphere no.9             & & 8541.4 g\\
  \quad Mass of sphere no.10            & & 8540.5 g\\
  \quad Mass of sphere no.12            & & 8541.7 g\\
  \quad Diameter of sphere no.7         & & 127.0 mm\\
  \quad Diameter of sphere no.9         & & 127.0 mm\\
  \quad Diameter of sphere no.10        & & 127.0 mm\\
  \quad Diameter of sphere no.12        & & 127.0 mm\\
  \quad Horizontal distance of GCs 7-9       & &342.3 mm\\
  \quad Horizontal distance of GCs 10-12     & &342.3 mm\\
  \quad Vertical distance of GCs 7-9         & &139.8 mm\\
  \quad Vertical distance of GCs 10-12       & &139.8 mm\\

  Relative positions in the TOS  experiment:   &\\
  \quad Centric height of pendulum                  &46.7 mm &\\
  \quad Centric height of sphere no.2               &46.7 mm &\\
 \quad  Centric height of sphere no.4               &46.7 mm &\\
  \quad Position of fiber in $X$ axis               &19 $\mu$m &\\
  \quad Position of fiber in $Y$ axis               &11 $\mu$m &\\
\hline
\hline
\end{tabular}
\end{table}

In order to calculate the terms $\Delta C_{g}$ and $\Delta C_{\text{sf}}(m_{\phi},\alpha_{s},\alpha_{p})$, we defines the laboratory coordinate system $(X,Y,Z)$ with the origin $O$ at the center-of-mass (CM) of the pendulum system. The $X$ axis points from the CM of source mass no.2 to CM of no.4. The $Z$ axis points along the torsion fiber, and $Y$ axis is defined by the right-handed relation. The pendulum's attitude is described by the angles $\theta_{X}$ and $\theta_{Y}$ about $X$ axis and $Y$ axis in the counterclockwise direction. The more detailed information can be found in the Ref. \cite{PhysRevD.82.022001}. In the laboratory coordinate system $(X,Y,Z)$, the Newtonian coupling coefficient $C_{g}$ is given by
\begin{equation}\label{dm3n}
  C_{g}=-M\int N(x,y,z,X_{2},Y_{2},Z_{2},\rho_{p},\theta)|_{\theta=0}dxdydz
\end{equation}
with
\begin{equation}\label{dm3nn}
  N(x,y,z,X_{2},Y_{2},Z_{2},\rho_{p},\theta)=\frac{\partial^{2}}{\partial^{2}\theta}
  \left[\frac{\rho_{p}}{\sqrt{(x-X_{2})^{2}+(y-Y_{2})^{2}+(z-Z_{2})^{2}}}\right],
\end{equation}
and the coupling coefficient $C_{\text{sf}}(m_{\phi},\alpha_{s},\alpha_{p})$ is similarly expressed as
\begin{equation}\label{dm3d}
  C_{\text{sf}}(m_{\phi},\alpha_{s},\alpha_{p})=-\alpha_{s}MI(m_{\phi}R)\int  M(x,y,z,X_{2},Y_{2},Z_{2},\rho_{p},\theta,m_{\phi},\alpha_{p})|_{\theta=0}dxdydz
\end{equation}
with
\begin{equation}\label{dm3nn1}
  M(x,y,z,X_{2},Y_{2},Z_{2},\rho_{p},\theta,m_{\phi},\alpha_{p})=\frac{\partial^{2}}{\partial^{2}\theta}
  \left[\frac{\alpha_{p}\rho_{p}e^{-m_{\phi}\sqrt{(x-X_{2})^{2}+(y-Y_{2})^{2}+(z-Z_{2})^{2}}}}
  {\sqrt{(x-X_{2})^{2}+(y-Y_{2})^{2}+(z-Z_{2})^{2}}}\right],
\end{equation}
where $M$ is the mass of sphere 2, $(X_{2},Y_{2},Z_{2})$ is the coordinate of the center of source mass, $(x,y,z)$ is coordinate of point mass of pendulum.

For the convenience of calculation, we introduce the pendulum coordinate system $(X_{0},Y_{0},Z_{0})$. Its origin $O_{0}$ is at the geometric center (GC) of the pendulum and the coordinate of $O_{0}$ at the laboratory coordinate system is $\delta X,\delta Y,\delta Z$. The $X_{0}$ axis is along the pendulum-length direction and the $Z_{0}$ points up along the height direction of the pendulum. The $Y_{0}$ axis is defined by the right-handed relationship. When the pendulum is rotated about the torsion fiber by an angle $\theta$, the coordinate transformation between two coordinate systems is given by
\begin{eqnarray}
\left(
  \begin{array}{c}
    X_{0} \\
    Y_{0} \\
    Z_{0}
  \end{array}
  \right)&=&
  {\cal R}\left(
  \begin{array}{c}
    X -\delta X\\
    Y -\delta Y\\
    Z- \delta Z
  \end{array}
  \right) \,,
\end{eqnarray}

where the transformation matrix is
\begin{eqnarray}
{\cal R} &=&
  \left(
  \begin{array}{ccc}
    \cos\beta\cos\theta & \cos\beta\sin\theta & -\sin\beta \\
    \sin\alpha\sin\beta\cos\theta-\cos\alpha\sin\theta & \sin\alpha\sin\beta\sin\theta+\cos\alpha\cos\theta & \sin\alpha\cos\beta \\
    \cos\alpha\sin\beta\cos\theta+\sin\alpha\sin\theta & \cos\alpha\sin\beta\sin\theta-\sin\alpha\cos\theta & \cos\alpha\cos\beta
  \end{array}
  \right) \,.
\end{eqnarray}

Then, we can transform  $N(x,y,z,\rho_{p},\theta)$ and $M(x,y,z,\rho_{p},\theta,m_{\phi},\alpha_{p})$ into $N(x_{0},y_{0},z_{0},\rho_{p},\theta)$ and $M(x_{0},y_{0},z_{0},\rho_{p},\theta,m_{\phi},\alpha_{p})$ by using the transformation matrix $\cal R$. The Newtonian coupling coefficient $C_{g}$ can be calculated by
\begin{equation}\label{dm5n}
  C_{g}=-M\int^{L/2}_{-L/2} dx_{0} \int^{W/2}_{-W/2}dy_{0} \int^{H/2}_{-H/2} dz_{0}
   N(x_{0},y_{0},z_{0},\rho_{p},\theta)|_{\theta=0}\cdot \left|\frac{\partial (X,Y,Z)}{\partial(X_{0},Y_{0},Z_{0})}\right|,
\end{equation}
where $L$, $W$, and $H$ represent the length, width, and height of the pendulum, respectively. Similarly, the {coupling parameter} $C_{\text{sf}}$ becomes
\begin{equation}\label{dm5d}
  C_{\text{sf}}=-\alpha_{s}MI(m_{\phi}R) \int^{L/2}_{-L/2} dx_{0} \int^{W/2}_{-W/2}dy_{0} \int^{H/2}_{-H/2} dz_{0}
    M(x_{0},y_{0},z_{0},\rho_{p},\theta,m_{\phi},\alpha_{p})|_{\theta=0}\cdot \left|\frac{\partial (X,Y,Z)}{\partial(X_{0},Y_{0},Z_{0})}\right|.
\end{equation}

Finally, the Newtonian term $\Delta C_{g}$ is given by $\Delta C_{g}=\sum_{i}\left(  C_{gn,i}- C_{gf,i}\right)$, where the sum is over all the parts of the pendulum system. And {scalar field} term $\Delta C_{\text{sf}}$ is also given by $\Delta C_{\text{sf}}=\sum_{i}\left(  C_{\text{sf}n,i}- C_{\text{sf}f,i}\right)$.
All the parameters are obtained by the numerical integration over all parts of the pendulum system. The main parameters are given in Table. \ref{tab:tos}. The Newtonian coefficient $\Delta C_{g}/I$ of the TOS-I-Fiber 3 is obtained $24911.71(22)$ kg$\cdot$m$^{-3}$ \cite{li2018measurements}.

In the AAF experiments, the signal frequency $\omega_{s}$ is at $2\omega_{d}$ with $\omega_{d}$ the constant difference between the angular velocities of the two turntables. Letting $m=2$, the Newtonian term becomes
\begin{equation}\label{adm8}
   D_{g}=\frac{-16\pi}{I} \sum^{\infty}_{l=2}\frac{1}{2l+1}m q_{lm} Q_{lm},
\end{equation}
and {scalar field} term is
\begin{equation}\label{adm9}
   D_{\text{sf}}(m_{\phi},\alpha_{p},\alpha_{s})=\frac{-16\pi}{I} \sum_{l=2}^{\infty}\frac{1}{2l+1}mq^{\text{sf}}_{lm}Q^{\text{sf}}_{lm}.
\end{equation}

In order to calculate Eqs.(\ref{adm8}) and (\ref{adm9}), we use the same method as the TOS method. The integral functions should also be calculated in the pendulum coordinate system. Since the twist angle of the AAF fiber is about zero in the AAF experiment, the transformation matrix between the laboratory and pendulum coordinate systems is
\begin{eqnarray}
{\cal R} &=&
  \left(
  \begin{array}{ccc}
    \cos\beta& 0& -\sin\beta \\
    \sin\alpha\sin\beta &\cos\alpha & \sin\alpha\cos\beta\\
    \cos\alpha\sin\beta\ & -\sin\alpha & \cos\alpha\cos\beta
  \end{array}
  \right) \,.
\end{eqnarray}
Using this transformation matrix, the terms $D_{g}$, $D_{\text{sf}}(m_{\phi},\alpha_{p},\alpha_{s})$ and the corresponding uncertainties can be given by the numerical integration over all parts of the pendulum system. The main parameters of the AAF experiment are listed in Table. \ref{tab:tos}. The Newtonian coefficient of the AAF-III is obtained as 6926.334(75) kg$\cdot$m$^{-3}$ \cite{li2018measurements}.

\section{The calculation on {scalar-field} parameters for the HUST-20 torsion-pendulum experiment}\label{appe2}

\begin{table}[!t]
\caption{\label{tab:hust20} The main experimental parameters of the HUST-20 experiment \cite{PhysRevLett.124.051301}.}
\newcommand{\tabincell}[2]{\begin{tabular}{@{}#1@{}}#2\end{tabular}}
\begin{tabular}{lcccc}
\hline
\hline
\tabincell{l}
Experimental parameters   \qquad\qquad\qquad\qquad\qquad \,\,\,\,\,\,\,\, \,\,\,\,\,\,\,\,&HUST-20 values     \\
\hline
  I-shaped Pendulum:              &   \\
  \quad Tungsten test masses                 & $14.6\times0.2\times12.0$ mm$^{3}$      \\
  \quad Compensation masses                  & $14.6\times0.3\times12.0$ mm$^{3}$    \\
  Attractor:              &   \\
  \quad Tungsten source masses                 & $17.6\times0.2\times11.4$ mm$^{3}$    \\
  \quad Compensation masses                   & $17.6\times0.2\times11.4$ mm$^{3}$     \\
\hline
\hline
\end{tabular}
\end{table}

\begin{figure}
\includegraphics[width=0.5\textwidth]{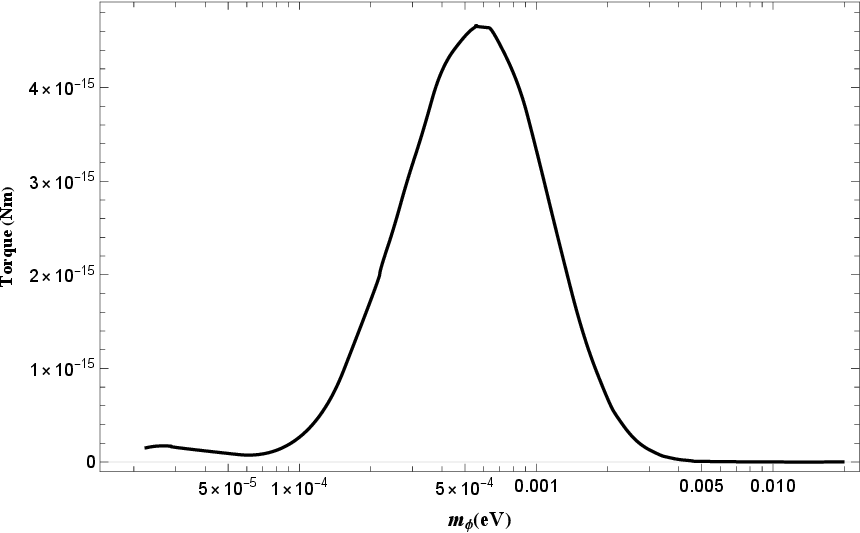}
\caption{\label{fig:isl} The calculated {scalar-field} torque for the separation 210 $\mu$m. The parameter $\alpha_{W}$ is assumed to be 1. }
\end{figure}

The schematic drawing of the experimental setup (HUST-20) is shown in Fig.1 of Ref. \cite{PhysRevLett.124.051301}. The main experimental parameters are listed in Table. \ref{tab:hust20}, more detailed information can be found in Ref. \cite{PhysRevLett.124.051301}. The 8$\omega_{d}$ measured torque at 210, 230, 295, and 1095 $\mu$m can be split into four data subsets, (a), (b), (c), and (d). For these four subsets, the mean values of the quadrature component and in-phase component are: (a) (-0.6,0.5)$\times10^{-17}$ Nm, (b) (-0.8,-0.6)$\times10^{-17}$ Nm, (c) (0.2,-0.5)$\times10^{-17}$ Nm, (d) (-0.6,-0.8)$\times10^{-17}$ Nm.
Using the experimental parameters, {the torques induced by the scalar fields} are obtained by the numerical integration
\begin{equation}\label{dd1}
  \tau_{\text{sf}}(m_{\phi},d_{a},\varphi_{n})=\int_{V_{i}}\int_{V_{j}}\frac{\partial}{\partial\theta}\left(\frac{G\rho_{i}\rho_{j}}{r}\alpha_{i}\alpha_{j}e^{-m_{\phi}r}\right)dx^{i}dy^{i}dz^{i}dx^{j}dy^{j}dz^{j}|_{\theta=0},
\end{equation}
where $\varphi_{n}$ is the rotation angle, and the integration is performed for all the parts of pendulum system. To calculate this torque, we calculate the integral function in the pendulum coordinate system. The different coordinate systems and the corresponding transformation matrix are similar to those in the Appendix.\ref{appe1}. For a rotation angle $\varphi_{n}$, the {scalar field} torque at $8\omega_{s}$ mainly comes from the component of sine
\begin{equation}\label{dd2}
 \tau_{\text{sf}} (8\omega_{s})=\frac{2}{N}\sum_{n}\tau_{\text{sf}}(\varphi_{n})\sin(8\varphi_{n}).
\end{equation}
where $\varphi_{n}=2\pi n/N$, $n=0,1,2...N-1$, and the component of cosine is given by a substitution from $\sin(8\varphi_{n})$ to $\cos(8\varphi_{n})$. Thus, the likelihood function is used to set constraints on the coupling coefficients. Assuming the parameter $\alpha_{W}=1$, for the experiment at separation 210 $\mu$m, Fig.\ref{fig:isl} shows the calculated {scalar-field} torques as a function of the field mass $m_{\phi}$.

\end{appendices}

%\bibliographystyle{elsarticle-num-names}
%\bibliographystyle{alpha}
%\bibliographystyle{apsrev4-1}
%\bibliographystyle{unsrt}

%\bibliography{dmandg}

%

\end{document}